\documentclass[usenatbib,twocolumn]{mn2e}
\usepackage{amsmath}
\usepackage{graphicx}
\usepackage{enumerate}
\usepackage{subfigure}
\usepackage[T1]{fontenc} 
\usepackage{aecompl}
\usepackage{verbatim}

\title{Baryonic effects on weak-lensing two-point statistics and its cosmological implications}
\author[I. Mohammed et al.] {Irshad Mohammed\thanks{irshad@physik.uzh.ch}$^{1,}$$^2$,
Davide Martizzi$^3$, Romain Teyssier$^2$ and Adam Amara$^4$\\
$^1${Physik-Institut, University of Zurich, Winterthurerstrasse 190,
  8057 Zurich, Switzerland} \\
$^2${Institute for Computational Science, University of Zurich,
  Winterthurerstrasse 190, 8057 Zurich, Switzerland} \\
$^3${Department of Astronomy, University of California, Berkeley, CA 94720-3411, USA}\\
$^4${Institute for Astronomy, Department of Physics, ETH Zurich, Wolfgang-Pauli-Strasse 27, 8093, Zurich, Switzerland}
}
\begin{document}
\maketitle
%===============================================================================
\begin{abstract}
We develop an extension of \textit{the Halo Model} that describes analytically the corrections to the matter power spectrum due to the physics of baryons. We extend these corrections to the weak-lensing shear angular power spectrum. Within each halo, our baryonic model accounts for: 1) a central galaxy, the major stellar component whose properties are derived from abundance matching techniques; 2) a hot plasma in hydrostatic equilibrium and 3) an adiabatically-contracted dark matter component. This analytic approach allows us to compare our model to the dark-matter-only case. Our basic assumptions are tested against the hydrodynamical simulations of Martizzi et. al. (2014), with which a remarkable agreement is found. Our baryonic model has only one free parameter, $M_{\rm crit}$, the critical halo mass that marks the transition between feedback-dominated halos, mostly devoid of gas, and gas rich halos, in which AGN feedback effects become weaker.  We explore the entire cosmological parameter space, using the angular power spectrum in three redshift bins as the observable, assuming a Euclid-like survey. We derive the corresponding constraints on the cosmological parameters, as well as the possible bias introduced by neglecting the effects of baryonic physics.  We find that, up to $\ell_{max}$=4000, baryonic physics plays very little role in the cosmological parameters estimation. However, if one goes up to $\ell_{max}$=8000, the  marginalized errors on the cosmological parameters can be significantly reduced, but neglecting baryonic physics can lead to bias in the recovered cosmological parameters  up to 10$\sigma$. These biases are removed if one takes into account the main baryonic parameter, $M_{\rm crit}$, which can also be determined up to 1-2\%, along with the other cosmological parameters.  
\end{abstract}
%===============================================================================
\begin{keywords}
Gravitational lensing: weak, 
methods: analytical,
galaxies: halos,
(cosmology:) cosmological parameters.
\end{keywords}
%===============================================================================
\section{Introduction}
\label{sec:intro}

The bending of light due to the presence of structures in its path is one very significant method to study the distribution of matter in the universe. The deflection is independent of the nature of the intervening matter, if it is dark or baryonic, and hence, this phenomenon, referred to as gravitational lensing, provides a unique tool to map the dark side of the universe. Under controlled systematics of the experiment, weak gravitational lensing, where the deflection of light rays are not significant enough to observe multiple images of the source but strong enough to deform the shape of the source, is a very powerful probe to study the nature of dark energy \citep{2006astro.ph..9591A}. The future sky surveys, like Euclid \citep{2011arXiv1110.3193L, 2009ExA....23...17R, 2009ExA....23...39C}, are expected to provide maps of the sky with un-precedented accuracy and high resolution like never before \citep{2013LRR....16....6A}. It is an opportunity  to employ the advantage of such high quality data to answer the most important questions in cosmology - the energy content of the universe, its dynamics, its evolution and the formation of structure. Weak gravitational lensing can be used as an ideal tool for such high quality data and can deliver, with sub-percent level accuracy, measurements of the main cosmological parameters.

The deformation of the shape of the observed galaxies due to the intervening matter is referred to as {\it shear}. This signal is very small, nearly 1$\%$ of the intrinsic ellipticity of the source galaxies, but can be measured statistically under the assumption that the intrinsic ellipticity of the background galaxies do not have a preferred direction. There are a number of interpretation of the two-point shear statistics based on dark matter only (collision-less) simulations which is a good approximation in the linear regime. However, at non-linear scales baryonic physics becomes important and can introduce a bias of 5 to 20 percent in the interpretation of the measurements, which in turn can introduce a bias in the cosmological constraints. So, in the era of precision cosmology, it is very important to quantify the effect of baryonic physics in the two-point shear statistics or the power spectrum.

Baryons account for nearly 20\% of the matter content of the universe. Its distribution depends on the dark matter potential well, AGN feedback, supernovae, structure formation history and radiative cooling. Further baryonic distribution affects the matter power spectrum at small scales, which to the extension, affects the two point shear statistics.  The effect of baryons on several statistics relevant for cosmology has been already studied by various authors. For instance, \cite{2009MNRAS.394L..11S,2012MNRAS.423.2279C,2014MNRAS.440.2290M} and \cite{2014MNRAS.439.2485C} focused on the effects on the halo mass function. The effect of baryonic processes on the power spectrum and on the weak gravitational lensing shear signal has been studied too \citep{2004APh....22..211W, 2004ApJ...616L..75Z, 2006ApJ...640L.119J, 2008ApJ...672...19R, 2010MNRAS.405..525G, 2011MNRAS.417.2020S, 2011MNRAS.415.3649V, 2014ApJ...783..118R, 2014arXiv1407.0060M}.

In most of the previous works (see references above), the approach was based on simulations, which suffer from finite volume and finite resolution effects, are performed using only one 
cosmology and baryonic model. They however capture the non-linear physics of gravitational collapse and the associated baryonic effects. In this work, we employ the halo model, an analytical approach, to build two-point shear statistics with and without baryons. This allows one to recover various different realizations of any cosmological models. We also compare our results with simulations at various stages to validate our main assumptions. 

The outline of the paper is as follows: In section \ref{sec:theory}, we review the necessary concepts of the halo model and propose our baryonic model as a modification in the radial density profiles of the halos. We compare the model to simulations with AGN feedback models. We also review the modelling of shear power spectrum. We talk about the covariance matrix of the $C_{\ell}$, Gaussian and non-Gaussian parts. In section \ref{sec:comparison}, we make a comparison between the dark-matter-only model (DMO) and our baryonic model (BAR) and shows the behaviour of the baryonic correction as a function of our main AGN-feedback-parameter, $M_{\rm crit}$. We introduce our fiducial model and mock datasets to perform the likelihood analysis in section \ref{sec:fiducial}. In section \ref{sec:cosmology}, we talk about the cosmological implication of these baryonic corrections and the forecasts on the cosmological parameters, its accuracy and precision. Finally in section \ref{sec:discussion} we discuss the implications of  our results and propose possible strategies for future works.

%===============================================================================

\section{Theoretical model - a short review}
\label{sec:theory}

We employ an analytic approach to model the effects of baryonic physics on the matter power spectrum and to the extension, on the shear power spectrum. The model has two broad parts: $(i)$ the dark-matter-only model (DMO), and $(ii)$ the modified model with baryonic physics (BAR). These two approaches modify the density profile of dark matter halos. We used the halo model \citep{2000MNRAS.318..203S,2000MNRAS.318.1144P,2000ApJ...543..503M,2002PhR...372....1C} to construct the matter power spectrum based on the density profiles of halos of mass $M$ and at redshift $z$.

%----------------------------------------------------------------------

\subsection{The halo model}\label{sec:halomodel}

We employed the halo model \citep{1977ApJ...217..331M,2000MNRAS.318..203S,2000ApJ...543..503M,2000MNRAS.318.1144P,2002PhR...372....1C} approach to calculate the matter power spectrum given the density profile of the halos. The halo model assumes all the matter in the universe to be in spherical halos with mass defined by a threshold density as:

\begin{equation}
	M_{\bigtriangleup} = \dfrac{4}{3} \pi R_{\bigtriangleup}^3 \bigtriangleup \bar{\rho}_{m}
\end{equation}
\\
where $M_{\bigtriangleup}$ is the mass of the halo and $R_{\bigtriangleup}$ is the boundary where the density of the halo drop to $\bigtriangleup$ times the mean matter density of the Universe, $\bar{\rho}_{m}$. We use $\bigtriangleup=200$ throughout this paper, unless stated otherwise. We define the virial radius of the halo $R_{\rm vir}$ to be $R_{200}$.

In this framework, the matter power spectrum can be split into two parts:
\begin{equation}
	P(k) = P_{1h}(k) + P_{2h}(k),
\end{equation}
\\
where, the two terms on the right hand side correspond to 1-halo term, describing the correlation between dark matter particles within the halo and 2-halo term which describes the halo-halo correlation respectively. These terms are given by

\begin{equation}
	P_{1h} = \int d\nu (f_{dm}+f_{gas}(\nu)) f(\nu) \dfrac{M}{\rho} |u(k|\nu)|^2, 
\end{equation}
\begin{equation}
	P_{2h} = \left(f_0b_0 + \int d\nu (f_{dm}+f_{gas}(\nu)) f(\nu) u(k|\nu) b(\nu) \right)^2 P_{\rm lin}(k),
\end{equation}
\\
where, $M$ is the mass of the halo and $\nu = \delta_c/\sigma(M,z)$ with $\delta_c = 1.686$. The term $f(\nu)$ is the functional form of the mass function and we used the fitting formula from \cite{2008ApJ...688..709T}. The term $b(\nu)$ resembles the bias in the dark matter halos and we used the fitting formula in \citep{2010ApJ...724..878T}. To fulfill the underlying assumptions of the halo model, these two functional forms, $f_{\nu}$ and $b_{\nu}$ have to be expressed as in the following relations:
\begin{equation}
	\int_{0}^{\infty} f(\nu)d\nu = 1
\end{equation}
\begin{equation}
	\int_{0}^{\infty} f(\nu)b(\nu)d\nu = 1
\end{equation}
\\
However, assuming a lower mass cut corresponding to $\nu_{\rm min}$, we introduce new background factors $f_0$ and $b_0$ such that:

\begin{equation}
	f_0 + \int_{\nu_{\rm min}}^{\infty} (f_{dm}+f_{gas}(\nu)) f(\nu)d\nu = 1
\end{equation}
\begin{equation}
	f_0b_0 + \int_{\nu_{\rm min}}^{\infty} (f_{dm}+f_{gas}(\nu)) f(\nu)b(\nu)d\nu = 1
\end{equation}
\\
Additionally, the term $f_{dm}+f_{gas} = 1$ for simpler models like no feedback, but for more exotic models, like with AGN feedback or including other baryonic physics, this term may deviate from unity. This will be more useful as explained in section \ref{sec:baryonicmodel}

\begin{figure*}
\centering
\subfigure{\includegraphics[width=0.48\textwidth]{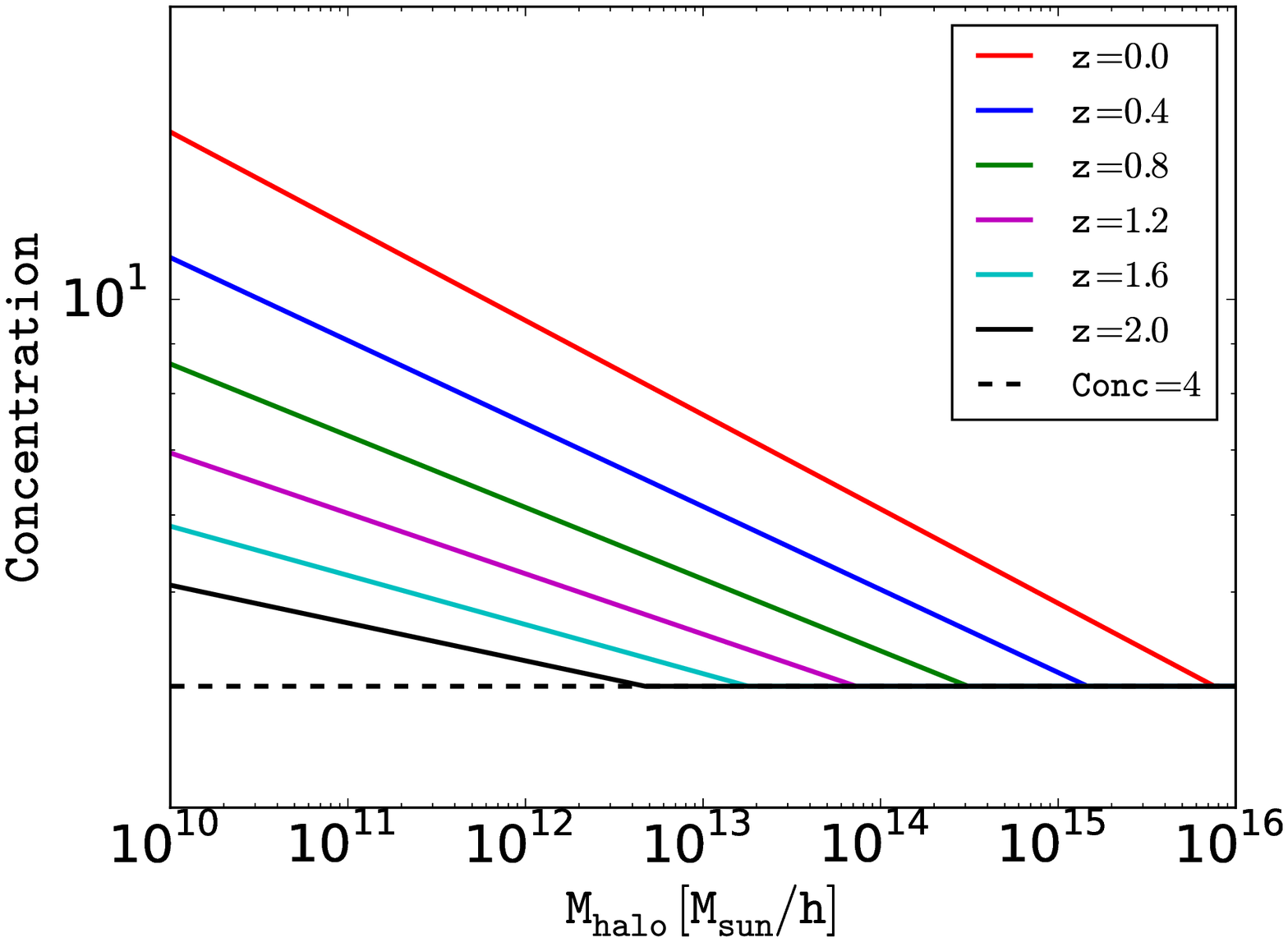}}
\subfigure{\includegraphics[width=0.48\textwidth]{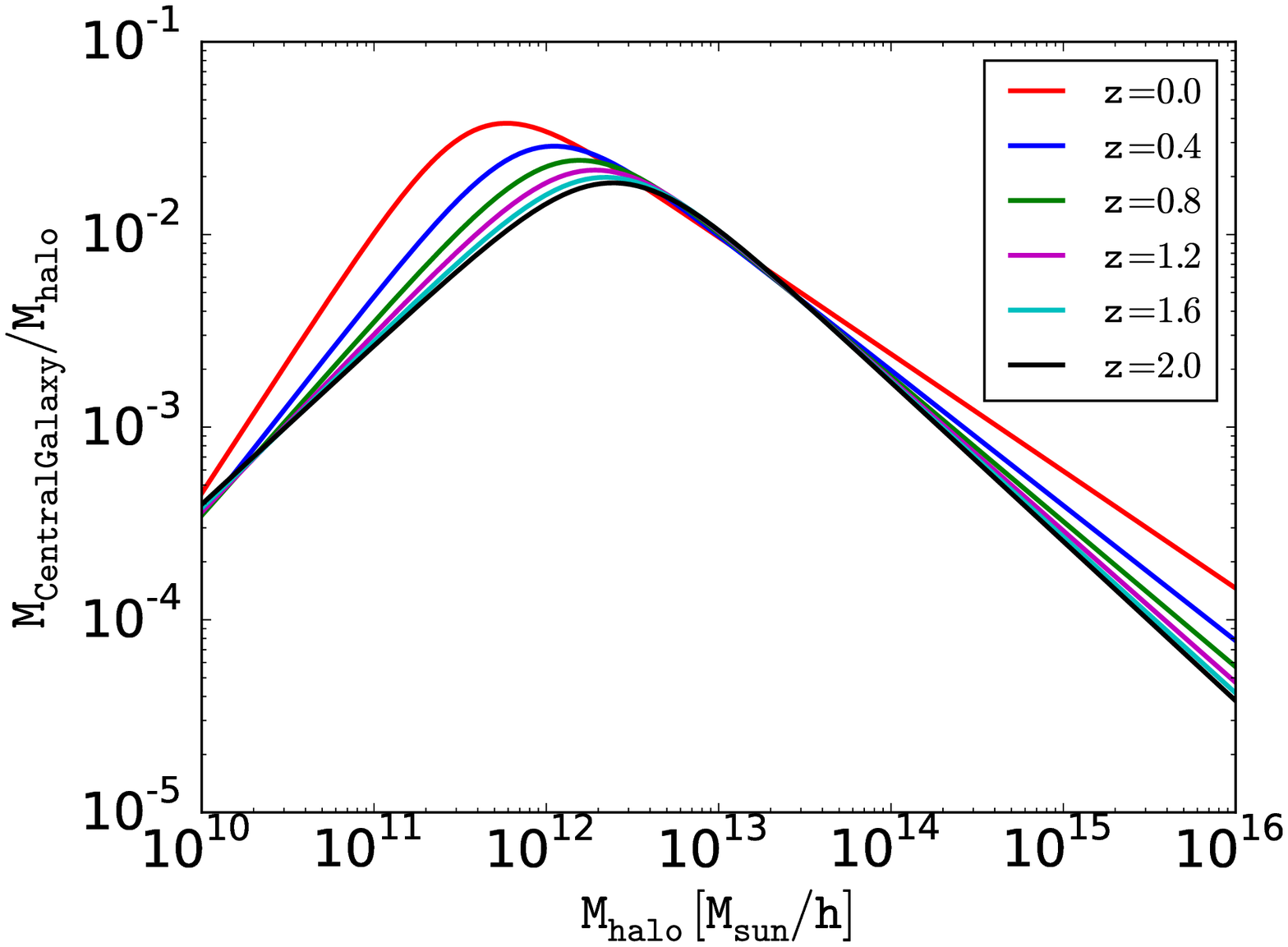}}
\subfigure{\includegraphics[width=0.48\textwidth]{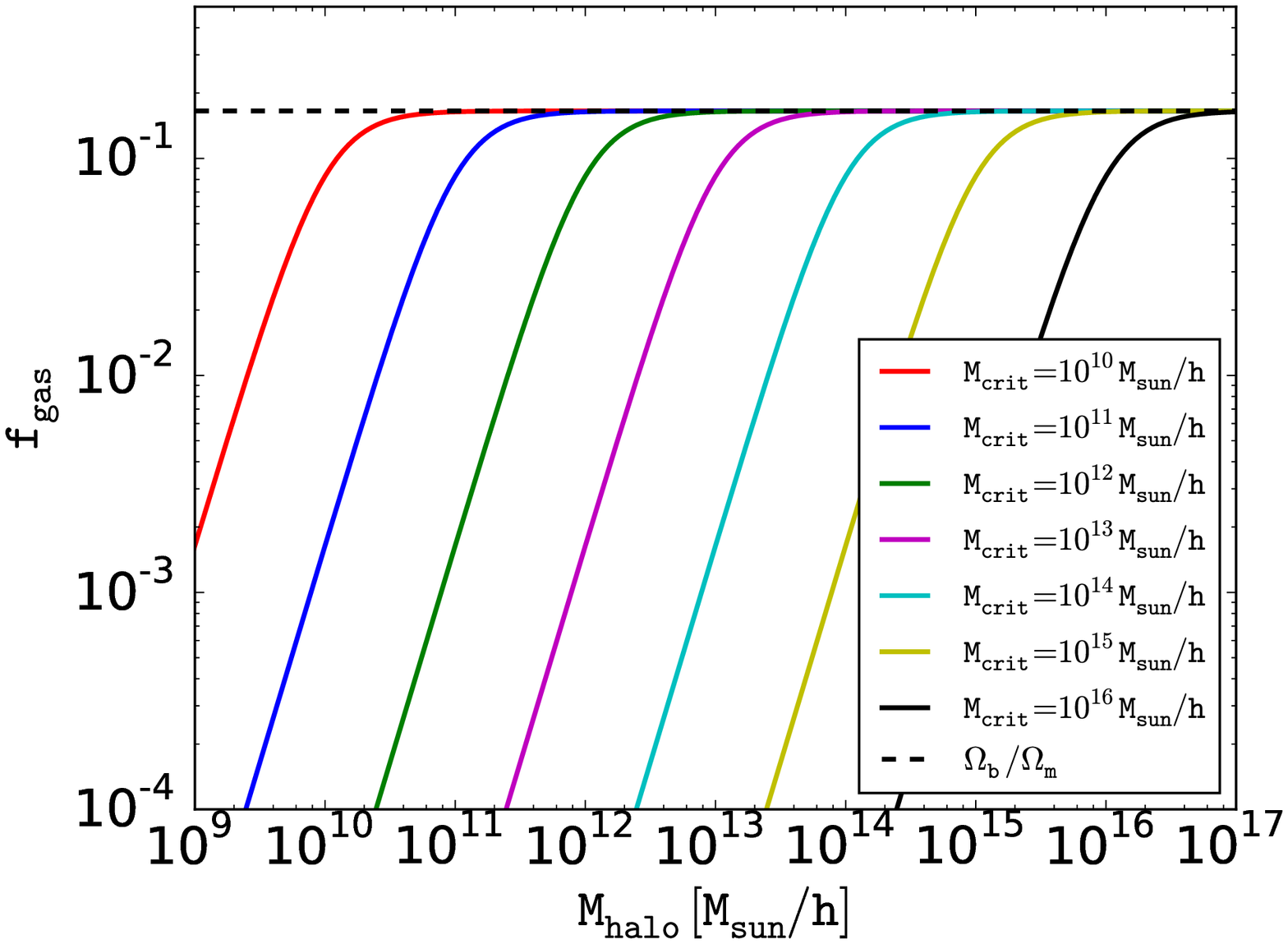}}
\subfigure{\includegraphics[width=0.48\textwidth]{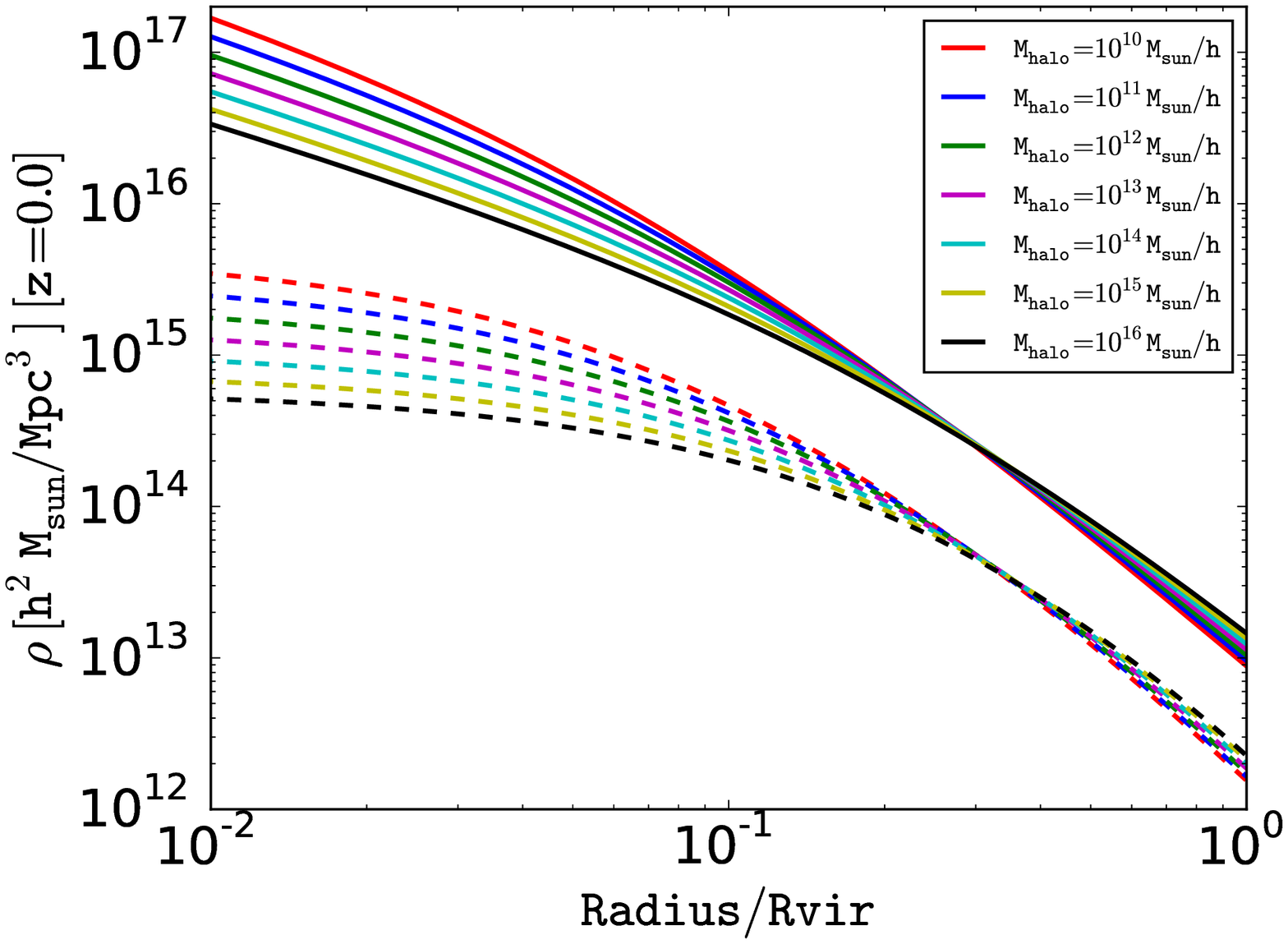}}
\caption{Top left: Concentration parameter as a function of halo mass for variable redshift. Top right: Mass of the central galaxy as a function of halo mass for variable redshift. Bottom left: Gas mass fraction as a function of halo mass for variable $M_{\rm crit}$. Bottom right: Density profile for NFW (solid lines) and intra-cluster gas (dashed lines) for different halo masses at redshift 0.}
\label{fig:halo}
\end{figure*}

We used the \cite{1998ApJ...496..605E, 1999ApJ...511....5E} transfer function calculations to account for the linear matter power spectrum term, $P_{lin}(k)$. The term $u(k|M)$ is the Fourier transform of the normalized density profile and is given by,

\begin{equation}
	u(k|M) = \dfrac{4 \pi}{M} \int_0^{R_{\rm vir}}dr\  r^2\  \rho(r|M)\ \dfrac{\sin(kr)}{kr}.
\end{equation}
\\
where, $\rho(r|M)$ is the density profile of the halo of mass $M$. The function $u(k|M)$ is normalised such that $u(k=0|M)=1$ .The dispersion of the smoothed density field, $\sigma(M,z)$, is given by,
\begin{equation}
	\sigma^2(M,z) = \dfrac{1}{2 \pi^2} \int P_{\rm lin}(k) k^2 |\tilde{W}(R,k)|^2 dk,
\end{equation}
\\
where, $\tilde{W}(R,k)$ is the Fourier transform of top-hat filtering function and given by,
\begin{equation}
	\tilde{W}(R,k) = 3 \dfrac{\textrm{sin}(kR) - kR \textrm{cos}(kR)}{(kR)^3}
\end{equation}

This framework of the halo model is applied to both DMO and BAR model which, differ in the halo density profiles and normalization of the mass function. The following two sections explains the corresponding profiles.

%----------------------------------------------------------------------
\subsection{Dark matter only}\label{sec:darkmatteronly}
We started with the radial density profile of dark matter halos given by the functional form:
\begin{equation}
 \rho(r|M) = \dfrac{\rho_s}{(r/R_s)^\alpha (1+r/R_s)^\beta},
\end{equation}
\\
where, $R_s$ is the characteristic radius given by the concentration parameter ($c$) and the virial radius of the halo ($r_{vir}$) as $c = R_{\rm vir}/R_s$. We used the two parameters $\alpha$ and $\beta$ to be 1 and 2 respectively, corresponding to the Navarro-Frenk-White (NFW) profile \citep{1997ApJ...490..493N}. The characteristic density $\rho_s$ which is strongly degenerate with $R_s$ and also proportional to the critical density of the Universe when the halo was formed. So, the NFW profile for dark matter halos is completely described by its concentration. 

The concentration parameter $c$ gives the information about the environment or the mean background density during the formation of the halo. A number of $N$-body simulations \citep{1997ApJ...490..493N, 1999MNRAS.310..527A, 2000ApJ...535...30J, 2001MNRAS.321..559B, 2001ApJ...554..114E, 2003ApJ...597L...9Z, 2007MNRAS.381.1450N, 2007MNRAS.378...55M, 2008MNRAS.390L..64D, 2008MNRAS.387..536G, 2014MNRAS.441.3359D} has prescribed various power laws between mass of the halo ($M$) and its concentration parameter $c$ at redshift $z$. We used the fitting formula given in \citep{2011MNRAS.411..584M}:

\begin{equation}
	\log(c) = a(z)\log(M_{\rm vir}/[h^{-1} M_{\bigodot}]) + b(z)
\end{equation}
\\
where,
\begin{equation}
	a(z) = \omega z - m
\end{equation}
\\
and
\begin{equation}
	b(z) = \dfrac{\alpha}{z+\gamma} + \dfrac{\beta}{(z+\gamma)^2}
\end{equation}
\\
The fitting parameters $\omega,\ m,\ \alpha,\ \beta\ {\rm and}\ \gamma$ are 0.029, 0.097, -110.001, 2469.720 and 16.885 respectively.
Figure \ref{fig:halo} (top-left panel) shows the behaviour of the concentration parameter as function of halo mass at different redshifts. There is an anti-correlation between the mass of the halo and its concentration. Also for a given halo mass, the concentration decreases with redshift. We limit the minimum concentration to 4 (dashed line in figure \ref{fig:halo} upper-left panel). This is because the higher mass halos did not reach there maximum formation efficiency redshift and will reach it in future. So, on an average, there concentration must not be less than a few. A very recent study from \cite{2014MNRAS.441.3359D} shows that this behaviour is consistent and the minimum concentration is very close to 4. 

%----------------------------------------------------------------------

\subsection{A baryonic model}\label{sec:baryonicmodel}
Our baryonic model accounts within each halo for: 1) a central galaxy, the major stellar component whose properties are derived from abundance matching techniques; 2) a hot plasma in hydrostatic equilibrium and 3) an adiabatically-contracted (AC) dark matter component. This analytic approach allows us to compare our model to the DMO case. Apart from the normalization of the mass function, there is only one term that is affected by these baryonic components and is the density profile of the halo, which no longer follows the NFW profile. We can write the modified NFW (BAR) profile as:
\begin{equation}
	\rho_{\rm BAR}(r|M) = f_{\rm dm}\rho_{\rm NFW}^{\rm AC}(r) + \rho_{\rm BCG}(r) + f_{\rm gas}(M)\rho_{\rm gas}(r),
\end{equation}
\\
we discuss each of these terms in more details.

%-------------------------------------------------------------------------

\subsubsection{Stellar component}\label{sec:stellar}
We used the fitting function from \cite{2013MNRAS.428.3121M} based on abundance matching to map the stellar mass of the central galaxy $M_{\rm CentralGalaxy}$ (BCG), which is the major component of stellar mass in a cluster, to the mass of the halo ($M_{\rm{halo}}$). Figure \ref{fig:halo} (top-right panel) shows the mapping between halo mass and stellar mass fraction associated to the central galaxy for a variety of redshifts. The relation has a positive slope for low mass halos, however, at about the size of the Milky way halo, the slope turns negative. At this peak, the central galaxy stellar mass contributes about 4-5 $\%$ of the total mass of the halo. Also this peak shifts to higher masses for higher redshifts but contributes lower fraction. 

The actual distribution of stellar mass in galaxy groups and clusters can be quite complex. The total stellar mass budget can be decomposed in 3 components: satellite galaxies, Brightest Cluster Galaxy (BCG, the massive elliptical galaxy dominating the cluster centre) and Intra-cluster Light (ICL, an extended stellar halo surrounding the BCG). The BCG and ICL represent $\sim 40$ \% of the mass in clusters, with this ratio decreasing with total cluster mass \citep{2007ApJ...666..147G}. However, BCG+ICL dominate the inner part of the cluster and constitute $\sim 70\%$ of the total stellar mass within 0.1 $R_{200}$. This fact is particularly relevant for computing the effect of baryon condensation on the dark matter profiles (see Subsection \ref{sec:adcon}). The BCG+ICL component is usually modelled using superimposition of fitting functions, typically multiple Sersic profiles. Given that we are not interested in detailed modelling of the stellar distribution, we consider a simplified model for the BCG+ICL.

we adopted a radial density profile for BCG, where the enclosed mass  goes linearly with the radius, 
\begin{equation}
	M_{\star}(<r) = M_{\rm CentralGalaxy} \dfrac{r}{2R_{1/2}}
\end{equation}
\\
this gives,
\begin{equation}
	\rho(r) = \dfrac{M_{\rm CentralGalaxy}}{8 \pi R_{1/2} r^2},\ r<2R_{1/2}
\end{equation}
\\
where, $R_{1/2}$ is the half mass radius. We use $R_{1/2} = 0.015 R_{\rm vir}$ which is a good fit to the observations \citep{2014arXiv1401.7329K}. We forced the density profile to drop exponentially after $2R_{1/2}$.

%We assume a Gaussian profile for the BCG which is normalized to its mass from the abundance matching relation ($M_{\rm{CentralGalaxy}}$) concentrated at the centre of the cluster.  Nearly all the mass of the BCG for most of the clusters is concentrated with a radius of 50 kpc. The concentration or the extent of this galaxy can be controlled with the Gaussian width, which is fixed to 5 kpc in this work. 

%-------------------------------------------------------------------------

\subsubsection{Intra-cluster plasma}\label{sec:gas}
\label{sec:icm}

The major component of the baryonic matter in a galaxy cluster is the hot intra-cluster gas. It is mainly ionized hydrogen at very high temperature and low density. This plasma radiates in X-rays and can safely be assumed to be in hydrostatic equilibrium. We assume this gas distribution in the halo according the hydrostatic equilibrium equations given in \cite{2013MNRAS.432.1947M},

\begin{equation}
	\rho(x) = \rho_0 \left[\dfrac{\ln(1+x)}{x}  \right]^{\dfrac{1}{\Gamma -1}}
\end{equation}
\\
where, $x$ is the distance from the centre of the halo in unit of scale radius $R_s$. The effective polytropic index $\Gamma$ is given by,

\begin{equation}
	\Gamma = 1+ \dfrac{(1+x_{eq})\ln(1+x_{eq}) - x_{eq}}{(1+3x_{eq})\ln(1+x_{eq})}
\end{equation}
\\
where, $x_{eq}=c/\sqrt{5}$. Figure \ref{fig:halo} (bottom-right in dashed lines) shows the density profile of the hot gas for variable halo masses at redshift 0 and also shows the comparison to the NFW profile (solid lines). For $x>x_{eq}$, the gas density profiles follows the NFW profile, however, it approaches a nearly constant values near the centre of the halo. 

The normalization of the gas density profile, $\rho_0$, is fixed by the gas fraction $f_{\rm{gas}}$. if we assume no feedback from the baryonic component of the halo, this number can be a constant, however, many hydrodynamical simulations \citep{2005MNRAS.356..107R, 2005MNRAS.360..892D, 2006Natur.442..539M, 2012MNRAS.421.3464P, 2013MNRAS.429.3068T, 2013MNRAS.432.1947M} shows signatures of the expulsion of gas from the halo. This expulsion is stronger in low mass halos than the high mass halos. So the low mass halos are generally deficit in this hot plasma component. Following the same physical motivation, we used the gas mass fraction of the halo to be the function of the mass of the halo following the parametric form:
\begin{equation}
	f_{\rm{gas}}(M_{\rm halo}) =  \dfrac{\Omega_b/\Omega_m}{1 + \left (\dfrac{M_{\rm{crit}}}{M_{\rm{halo}}} \right) ^{\beta}}
\end{equation}
\\
where, $M_{\rm{crit}}$ is a free parameter and $\beta$ is fixed to 2. This parameter controls the gas fraction in halos of different mass. A higher value for $M_{\rm{crit}}$ represents less gas in the halo up to higher halo masses. This parameter can also be interpreted as the control sequence for AGN feedback. Figure \ref{fig:halo} (bottom-left panel) shows the variation of $f_{\rm{gas}}$ with halo mass for variety of $M_{\rm{crit}}$. We chose $M_{\rm crit} = 10^{13} h^{-1} M_{\bigodot}$ as the most realistic model. In this case, all halos with mass lower than $\sim 2\times 10^{12} h^{-1} M_{\bigodot}$ have expelled all their gas to the background (outside the $R_{\rm vir}$) and all halos with mass larger than $\sim 2\times 10^{13} h^{-1} M_{\bigodot}$ have all their gas inside the halo. The intermediate mass halos have a very smooth transition from no gas to all gas inside the halo. This behaviour matches well with recent study from \cite{2014arXiv1409.8617S}. We studied this case in detail for all its cosmological implications at different scales. We also studied one optimistic\footnote{Optimistic in the sense of less AGN feedback that makes the baryonic corrections less troublesome} model, where the feedback is not as strong as in our realistic model, with $M_{\rm crit} = 10^{12} h^{-1} M_{\bigodot}$.

%-------------------------------------------------------------------------

\subsubsection{Adiabatic contraction}\label{sec:ac}
\label{sec:adcon}
In the DMO model, we adopted the NFW profile for the distribution of dark matter in the halo which is nearly scale-free and completely described by the concentration parameter. However, in the presence of baryons, the dark matter component follows NFW only in the outskirts of the halo, but in the very centre the dark matter profile becomes steeper and deviates from pure a NFW profile.  This is because the baryons, which are dominant in the centre of the halo, drag some extra matter from the surrounding towards the centre making the dark matter profile steeper towards the centre. The total distribution of matter is expected to dynamically respond to the condensation of baryons at the centre of the halo in a way that approximately conserves the value the adiabatic ``invariant'' $R\times M(R)$, where $R$ is the distance from the halo centre and $M(R)$ is the mass enclosed in a sphere of radius R \citep{1986ApJ...301...27B, 2004ApJ...616...16G}. We adopted a simplified model for this effect following the appendix of \cite{2011MNRAS.414..195T} where this adiabatic contraction (AC) of the dark matter profile is solely governed by the central galactic disk.

%----------------------------------------------------------------------

\subsection{Comparison with simulations}\label{sec:simulations}

\begin{figure*}
	\includegraphics[width=0.48\textwidth]{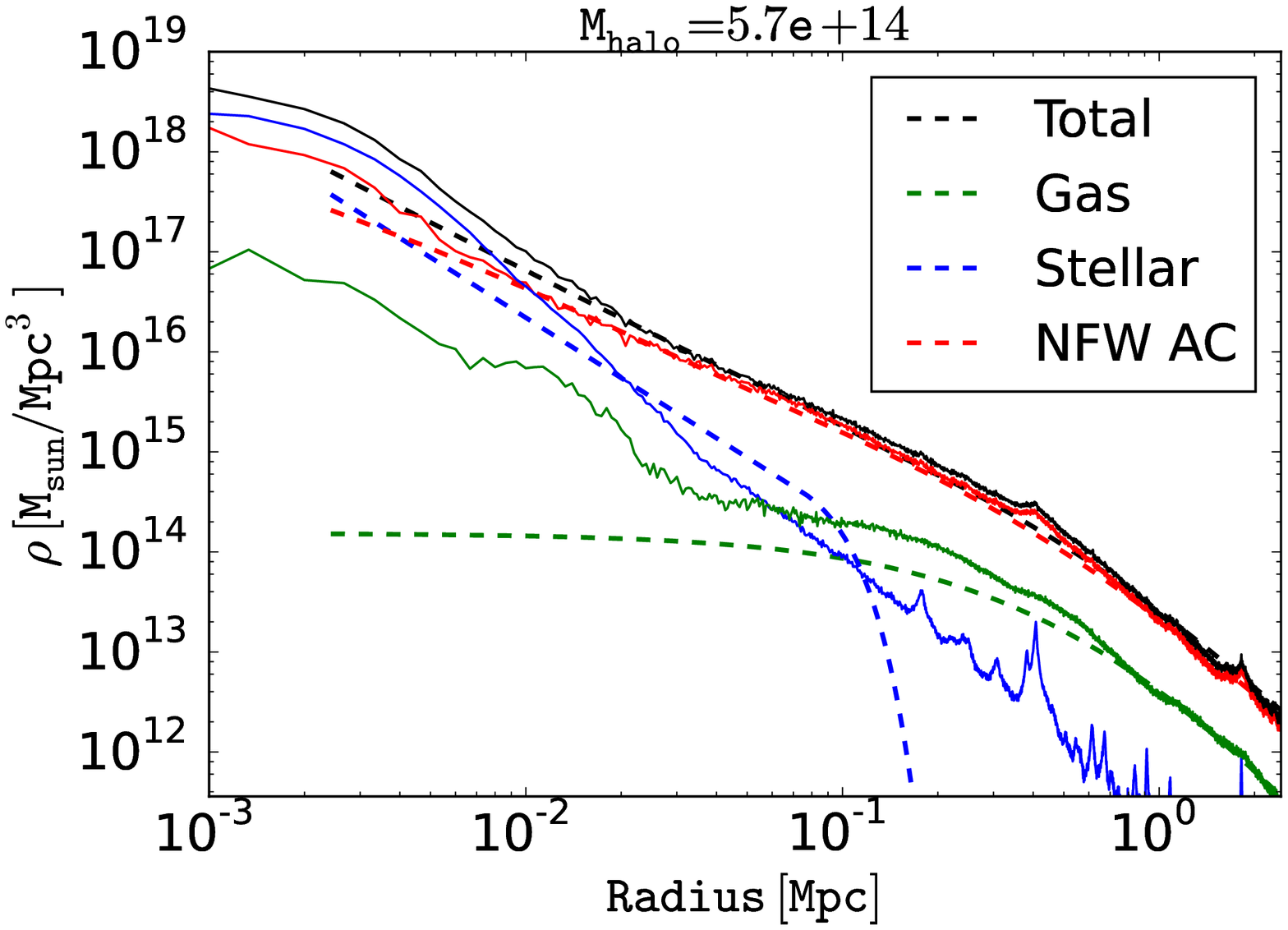}
	\includegraphics[width=0.48\textwidth]{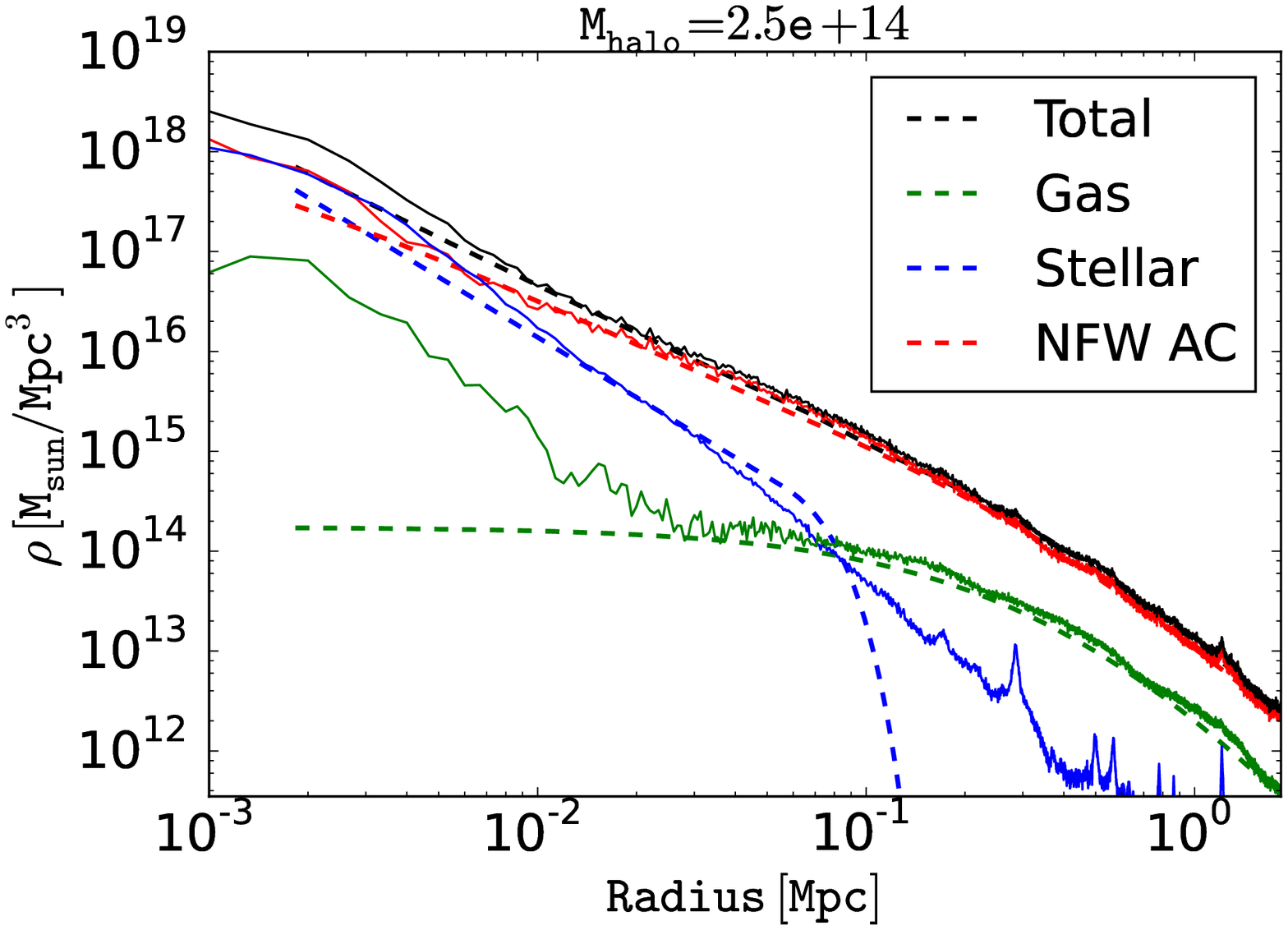}\\
	\includegraphics[width=0.48\textwidth]{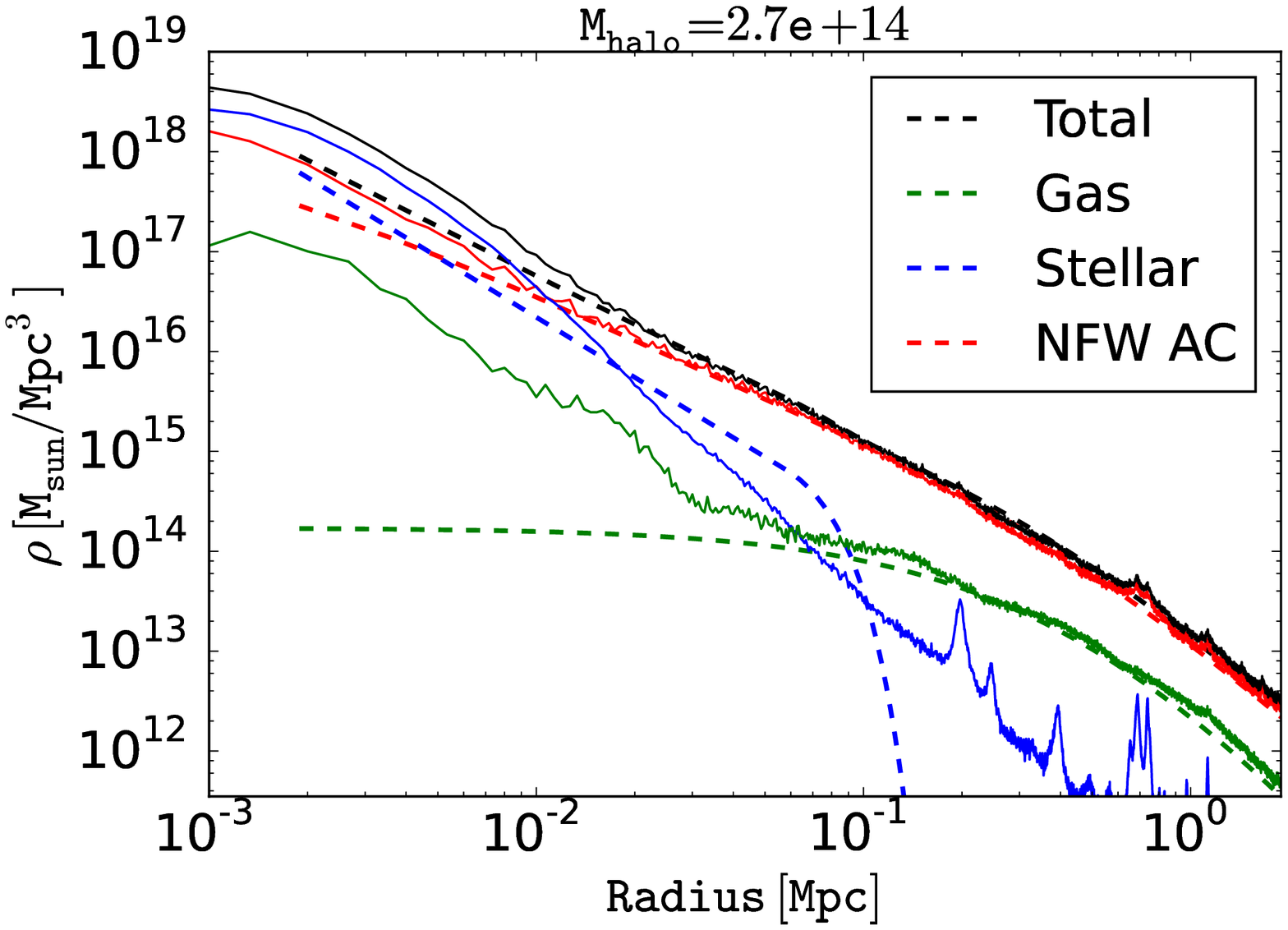}
	\includegraphics[width=0.48\textwidth]{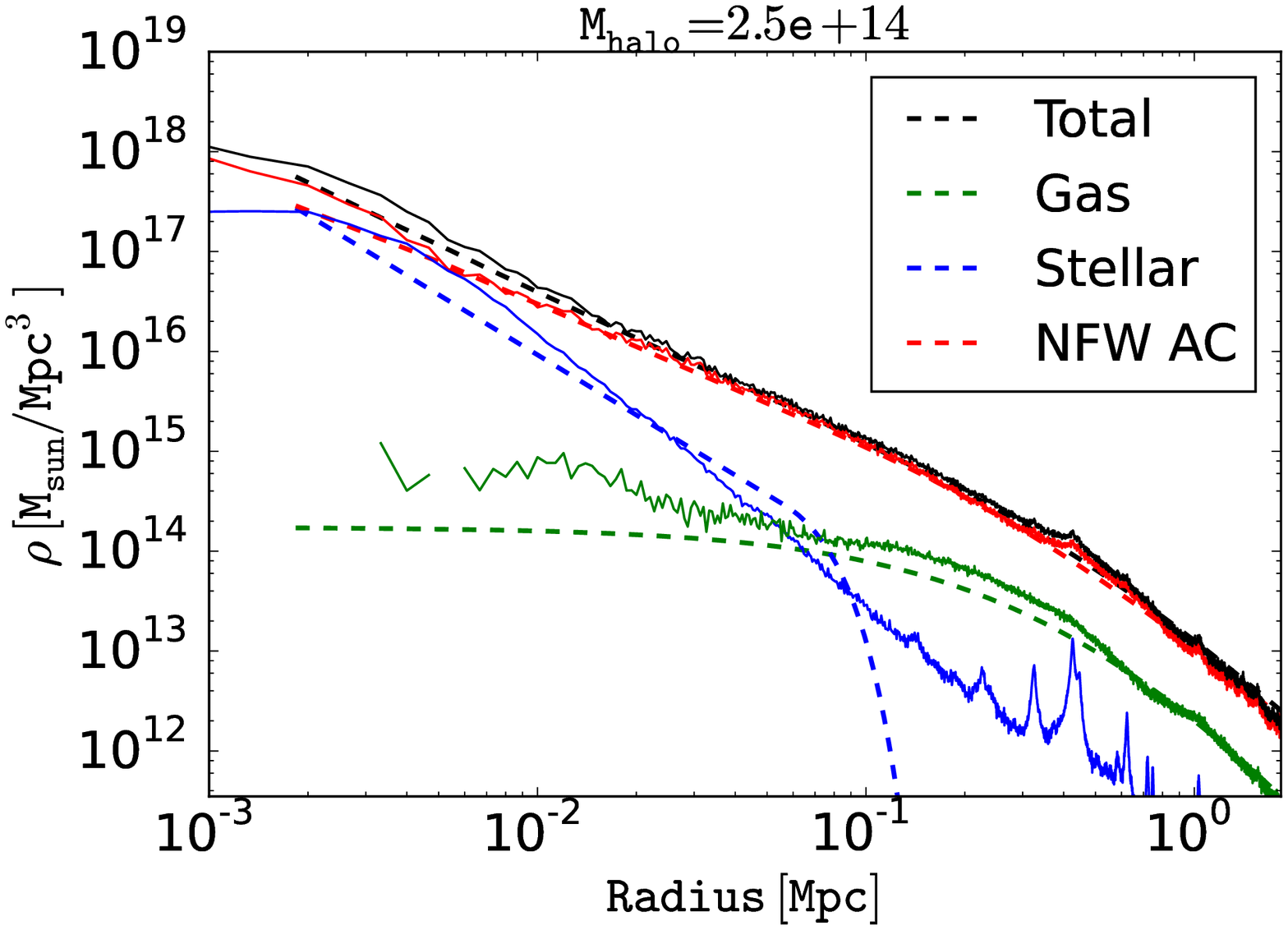}\\
	\includegraphics[width=0.48\textwidth]{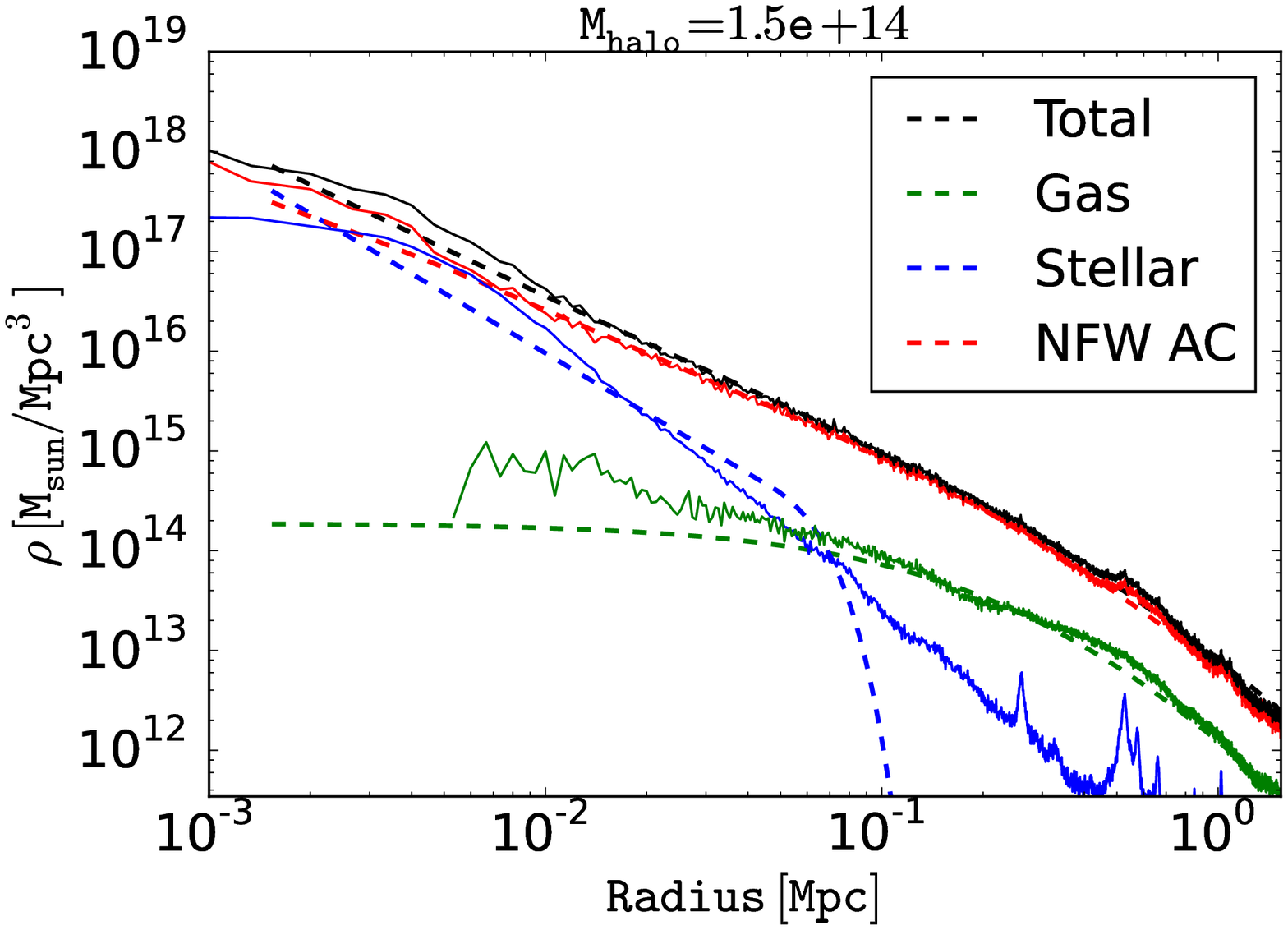}
	\includegraphics[width=0.48\textwidth]{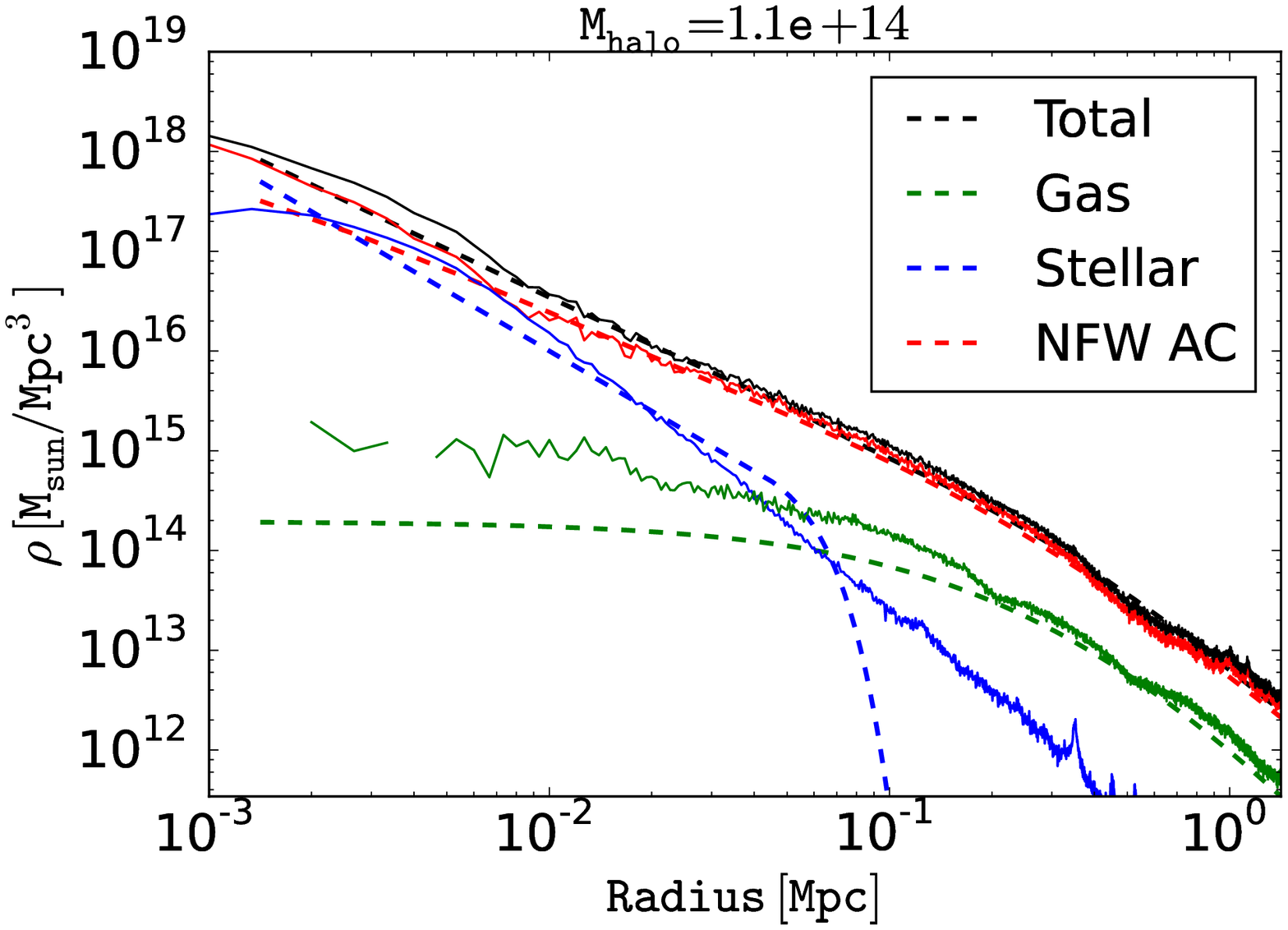}
	\caption{A comparison of our model density profiles (dashed lines) with hydrodynamical simulations of Martizzi et. al. 2014 (solid lines). There is a remarkable agreement, except at the very centre of the halo.}
	\label{fig:compare}
\end{figure*}

\begin{table}
\begin{center}
\begin{tabular}{|l|c|c|c|c|c|c|}
\hline
\hline
 Type & $H_0$ & $\sigma_{\rm 8}$ & $n_{\rm s} $ & $\Omega_\Lambda$ & $\Omega_{\rm m}$ & $\Omega_{\rm b}$ \\
\hline
\hline
DMO & 70.4 & 0.809 & 0.963 & 0.728 & 0.272 & - \\
BAR & 70.4 & 0.809 & 0.963 & 0.728 & 0.272 & 0.045 \\
\hline
\hline
\end{tabular}
\caption{Cosmological parameters adopted in our simulations. }\label{tab:cosm_par}
\end{center}
\end{table}

\begin{table}
\begin{center}
\begin{tabular}{|l|c|c|c|}
\hline
\hline
{\itshape Type} & $m_{\rm cdm}$&  $m_{\rm gas}$ & $\Delta x_{\rm min}$ \\
 & $[10^{8}$ M$_\odot$/h] & $[10^{7}$ M$_\odot$/h] & [kpc/h] \\
\hline
\hline      
 Original box & $ 15.5 $ & n.a. & $2.14$ \\
 DMO zoom-in & $1.94$ & n.a. & $1.07$ \\
 BAR zoom-in & $1.62$ & $3.22$ & $1.07$ \\
\hline
\hline
\end{tabular}
\end{center}
\caption{Mass resolution for dark matter particles, gas cells and star particles, and spatial resolution (in physical units) for our simulations. }\label{tab:mass_par}
\end{table}

We consider data from a set of cosmological re-simulations performed with the {\scshape ramses} code \cite{2002A&A...385..337T}. These simulations are part of a larger set recently used 
by \cite{2014MNRAS.440.2290M} to study the baryonic effects on the halo mass function. Thanks to the adaptive mesh refinement capability of the {\scshape ramses} code, the resolution 
achieved in these simulations is sufficient to study the properties of low redshift BCGs. 

In these calculations, the cosmological parameters are: matter density parameter $\Omega_{\rm m}=0.272$, cosmological constant density parameter $\Omega_\Lambda=0.728$, baryonic matter 
density parameter $\Omega_{\rm b}=0.045$, power spectrum normalization $\sigma_{\rm 8}=0.809$, primordial power spectrum index $n_{\rm s}=0.963 $ and Hubble constant $H_0=70.4$ km/s/Mpc 
(Table~\ref{tab:cosm_par}). We generated initial conditions for the simulations using the \cite{1998ApJ...496..605E} transfer function and the {\scshape grafic++} code\footnote{http://sourceforge.net/projects/grafic/}, based on the original {\scshape grafic} code \citep{2001ApJS..137....1B}. These simulations come in two flavours: DMO (dark matter only) 
which only follow the evolution of dark matter, BAR which include baryons and galaxy formation prescriptions. 

The technique we adopted to perform the zoom-ins is described in the following. First, we ran a dark matter only simulation with particle mass $m_{\rm cdm}=1.55\times 10^9$~M$_\odot$/h 
and box size $144$~Mpc/h. The initial level of refinement was $\ell=9$ ($512^3$), but as the simulation evolved more levels of refinement were allowed. At redshift $z=0$ the grid was 
refined down to a maximum level $\ell_{\rm max}=16$. Subsequently, we ran apply the AdaptaHOP algorithm \cite{2004MNRAS.352..376A} to identify the position and masses of dark 
matter halos. We selected 51 halos whose {\it total} masses lie $M_{\rm tot}>10^{14}$~M$_\odot$ and whose neighbouring halos do not have masses larger than $M/2$ within a spherical 
region of five times their virial radius. We determined that only 25 of these clusters are relaxed. High resolution initial conditions were extracted for each of the 51 halos and were used 
to run zoom-in re-simulations. Three different re-simulations per halo have been performed: (I) including dark matter and neglecting baryons, (II) including dark matter, baryons 
and stellar feedback, (III) including baryons, stellar feedback and AGN feedback. In this paper we focus on cases (I) and (III), labelled DMO and BAR, respectively.

In the DMO re-simulation, the dark matter particle mass is $m_{\rm cdm}=1.94\times 10^{8}$~M$_\odot$/h. In the BAR re-simulations, the dark matter particle mass is 
$m_{\rm cdm}=1.62\times 10^{8}$~M$_\odot$/h, while the baryon resolution element has a mass of $m_{\rm gas}=3.22\times 10^{7}$~M$_\odot$. The maximum refinement level was set to $\ell=17$, 
corresponding to a minimum cell size $\Delta x_{\rm min} = L/2^{\ell_{\rm max}}\simeq 1.07$ kpc/h. The grid was dynamically refined using a quasi-Lagrangian approach: 
when the dark matter or baryonic mass in a cell reaches 8 times the initial mass resolution, it is split into 8 children cells. Table~\ref{tab:mass_par} summarizes the particle mass and 
spatial resolution achieved in the simulations.

The physical prescription implemented in the code to perform the BAR simulations is here briefly described. In {\scshape ramses} gas dynamics is solved via a second-order unsplit Godunov 
scheme \citep{2002A&A...385..337T} based on different Riemann solvers (we adopted the HLLC solver) and the MinMod slope limiter. The gas is described by perfect gas equation of state 
(EOS) with polytropic index $\gamma=5/3$. Gas cooling is modelled with the \cite{1993ApJS...88..253S} cooling function which accounts for H, He and metals. Star formation and supernovae 
feedback ("delayed cooling" scheme, \cite{2006MNRAS.373.1074S}) and metal enrichment have been included in the calculations. AGN feedback has been included too, using a method inspired by 
the \cite{2009MNRAS.398...53B} model. In this scheme, super-massive black holes (SMBHs) are modeled as sink particles and AGN feedback is provided in form of thermal energy injected in a 
sphere surrounding each SMBH. More details about the AGN feedback scheme and about the tuning of the galaxy formation prescriptions can be found in \cite{2011MNRAS.414..195T} and \cite{2012MNRAS.422.3081M}.

Figure~\ref{fig:compare} shows the comparison between the dark matter, gas, stellar and total mass density profiles of 6 halos in the \cite{2014MNRAS.440.2290M} catalogue and the mass model described in Section \ref{sec:halomodel}. The model for the adiabatically contracted dark matter profile (red dashed lines) fits well the simulations down to scales $\sim 10$ kpc. The model for the Intra-cluster plasma (green dashed lines) fits well the results of the simulations down to scales $\sim 50$ kpc. The relation between mass of the central galaxy and that of the halo has a lot of scatter. So, to compare with simulations we use the stellar mass from the simulation itself for the given halo, which define the normalisation of our stellar model. The model (blue dashed lines) is a good fit to the results of the simulations except in the outskirts. This is expected since the data from the simulations include BCG, ICL and satellite galaxies. However, the model is constructed in such a way that the stellar mass expected from abundance matching is associated to the central regions of the halos. The overall result is that the model for the total mass (black dashed lines) provides an excellent match to the results of cosmological simulations down to a scale of $\sim 10$ kpc. Therefore we conclude that the mass model is good enough to be adopted for the purposes of this paper.

%===============================================================================

\subsection{From $P(k)$ to $C(\ell)$}
\label{sec:pk2cl}
In this section we develop the mapping from 3D matter power spectrum $P(k,z)$ to the 2D projected shear angular power spectrum $C_{\ell}$ following the theoretical framework explained in \cite{2009MNRAS.395.2065T}.

The distortion of the source shape due to weak gravitational lensing can be quantified with two quantities: shear $\gamma$ and convergence $\kappa$. The convergence $\kappa$ is the local isotropic part of the deformation matrix and can be expressed as:
\begin{equation}
	\kappa(\vec{\theta}) = \dfrac{1}{2} \vec{\bigtriangledown}.\vec{\alpha}(\vec{\theta})
\end{equation}
\\
where, $\alpha$ is the deflection angle. If we know the redshift of the source galaxies, additional information can be gained by dividing the sources in different redshift bins. This process is referred to as lensing tomography and is very useful to gain extra constraints on cosmology from the evolution of the weak lensing power spectrum \citep{1999ApJ...522L..21H, 2002PhRvD..65f3001H, 2004MNRAS.348..897T}. In cosmological context, the convergence field can be expressed as the weighted projection of the mass distribution integrated along the line of sight in the $i$th redshift bin, 

\begin{equation}
	\kappa_i(\vec{\theta}) =  \int_0^{\chi_H} g_i(\chi) \delta(\chi \vec{\theta},\chi)d\chi,
\end{equation}
\\
where, $\delta$ is the total 3 dimensional matter overdensity, $\chi$ is the comoving distance and $\chi_H$ is the comoving distance to the horizon. For a complete review see \cite{1999ARA&A..37..127M, 2001PhR...340..291B, 2006glsw.conf..269S}. The lensing weights $g_i(\chi)$ in the $i$th redshift bin with comoving distance range between $\chi_i$ and $\chi_{i+1}$ are given by:

\begin{equation}
g_i(\chi) = \begin{cases} \dfrac{g_0}{\bar{n}_i} \dfrac{\chi}{a(\chi)} \int_{\chi_i}^{\chi_{i+1}} n_s(\chi^{\prime})\dfrac{dz}{d\chi^{\prime}} \dfrac{(\chi^{\prime} - \chi) }{\chi^{\prime}} d\chi^{\prime}, & \chi \le \chi_{i+1} \\ 
			0, & \chi > \chi_{i+1} \end{cases}
\end{equation}
\\
where, $a(\chi)$ is the scale factor at comoving distance $\chi$. Also, 

\begin{equation}
	g_0 = \dfrac{3}{2} \dfrac{\Omega_m}{H_0^2}
\end{equation}
\\
and,
\begin{equation}
	\bar{n}_i = \int_{\chi_i}^{\chi_{i+1}} n_s(\chi(z)) \dfrac{dz}{d\chi^{\prime}} d\chi^{\prime}.
\end{equation}
\\
where, $n_s(\chi(z))$ is the distribution of sources in redshift. We assume a source distribution along the line of sight of the form:
\begin{equation}
	n_s(z) = n_0 \times 4z^2 \exp\left(-\dfrac{z}{z_0}   \right)
\end{equation}
\\
with $n_0 = 1.18 \times 10^{9} $ per unit steradian and $z_0$ is fixed such that the corresponding projected source density $n_g$ resembles the experiment, like Euclid etc.

\begin{equation}
	\int_0^{\infty} n_s(z)dz = \bar{n}_g.
\end{equation}
\\
For Euclid like survey, we choose $z_0$ such that $\bar{n}_g=50$ sources per arcmin$^{-2}$ \citep{2008ARNPS..58...99H}.

Finally the shear power spectrum between redshift bins $i$ and $j$ can be computed as:

\begin{equation}
	C_{ij}(\ell) =  \int_0^{\chi_H} \dfrac{g_i(\chi) g_j(\chi)}{ \chi^2} P\left(\dfrac{\ell}{\chi},\chi \right)d \chi
\end{equation}
\\
where, $P$ is the 3D matter power spectrum calculated using the halo model framework as described in section \ref{sec:halomodel}. Larger $\ell$ corresponds to the smaller scale and the large contribution of $C_{\ell}$ at higher $\ell$ comes from non-linear clustering. 

We divided the big cosmological volume into 3 redshift bins with boundaries: 0.01, 0.8, 1.5 and 4.0; so we calculated total 6 convergence cross-spectra (3 auto-spectra and 3 cross-spectra).

The auto-spectra is contaminated by the intrinsic ellipticity noise and assuming its distribution to be completely uncorrelated to different source galaxies, the observed power spectrum $C_{ij}^{\rm{obs}}(\ell)$ is given by,

\begin{equation}
	C_{ij}^{\rm{obs}}(\ell) = C_{ij}(\ell) + \delta_{ij}\dfrac{\sigma_{\epsilon}^2}{\bar{n}_i},
\end{equation}
\\
we choose $\sigma_{\epsilon}=0.33$ which is the RMS intrinsic ellipticity. The cross spectra is not contaminated by shot noise. 

The covariance matrix of $C_{\ell}$ has two contributions: Gaussian and non-Gaussian (NG). In this work we only consider the Gaussian contribution to the covariance matrix which is given by the following expression,
%\begin{equation}
\begin{align}
	{\rm Cov}_{ij,mn}(\ell,\ell^{\prime})& = \dfrac{\delta_{\ell\ell^{\prime}}}
										{\Delta\ell(2\ell+1) \rm{f_{sky}}} \times  \nonumber\\
										&\qquad  \left(C_{im}^{\rm{obs}}(\ell)C_{jn}^{\rm{obs}}(\ell)+
										C_{in}^{\rm{obs}}(\ell)C_{jm}^{\rm{obs}}(\ell) \right),
\end{align}
%\end{equation}
\\
where, $\Delta\ell$ is the bin width of the $\ell$ and $f_{sky}$ is the sky fraction for the targeted experiment. This term is dominated by cosmic variance for lower $\ell$ and shot noise for higher $\ell$, however, for large number of sources, as in case of Euclid, and larger size of bins $(\Delta \ell)$ towards higher end of $\ell$, the shot noise can be significantly reduced.

The NG contribution to the covariance matrix of $C_{\ell}$ is rather complicated to calculate. It gives the correlation between different $\ell$. At the matter power spectrum level, this term depends on the matter trispectrum. To compute the NG covariance to lensing, we need to integrate the trispectrum in redshift and angle on the sky and then compute this quantity for various $\ell$ and $\ell^{\prime}$. So this is a 4D calculation of trispectrum which is computationally very expensive. \cite{2012PhRvD..86h3504Y} shows that these NG correction to the covariance becomes significant for $\ell$ of few thousand and \cite{2002PhR...372....1C} shows that neglecting this will introduce the bias in the cosmological parameters up to 20 $\%$. In this work, we are not taking into account these corrections and we are doing our analysis for different $\ell_{max}$: 1000, 2000, 3000, 4000, 5000, 6000, 8000, 10000 and 20000. We will discuss more about the NG covariance in section \ref{sec:NG}. 
%----------------------------------------------------------------------

%===============================================================================

\section{Comparing BAR and DMO model}
\label{sec:comparison}

\begin{figure*}
	\includegraphics[width=0.48\textwidth]{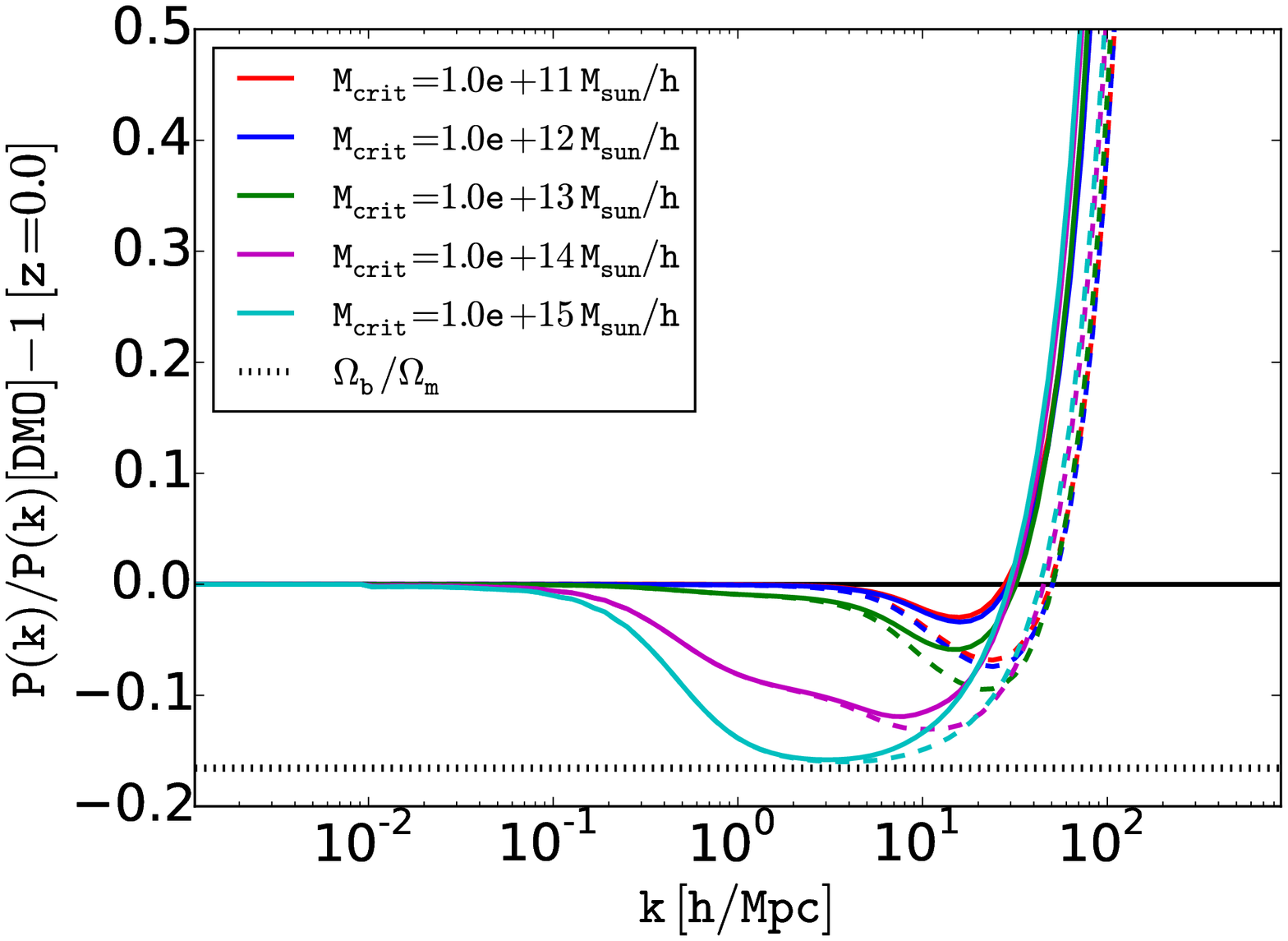}
	\includegraphics[width=0.48\textwidth]{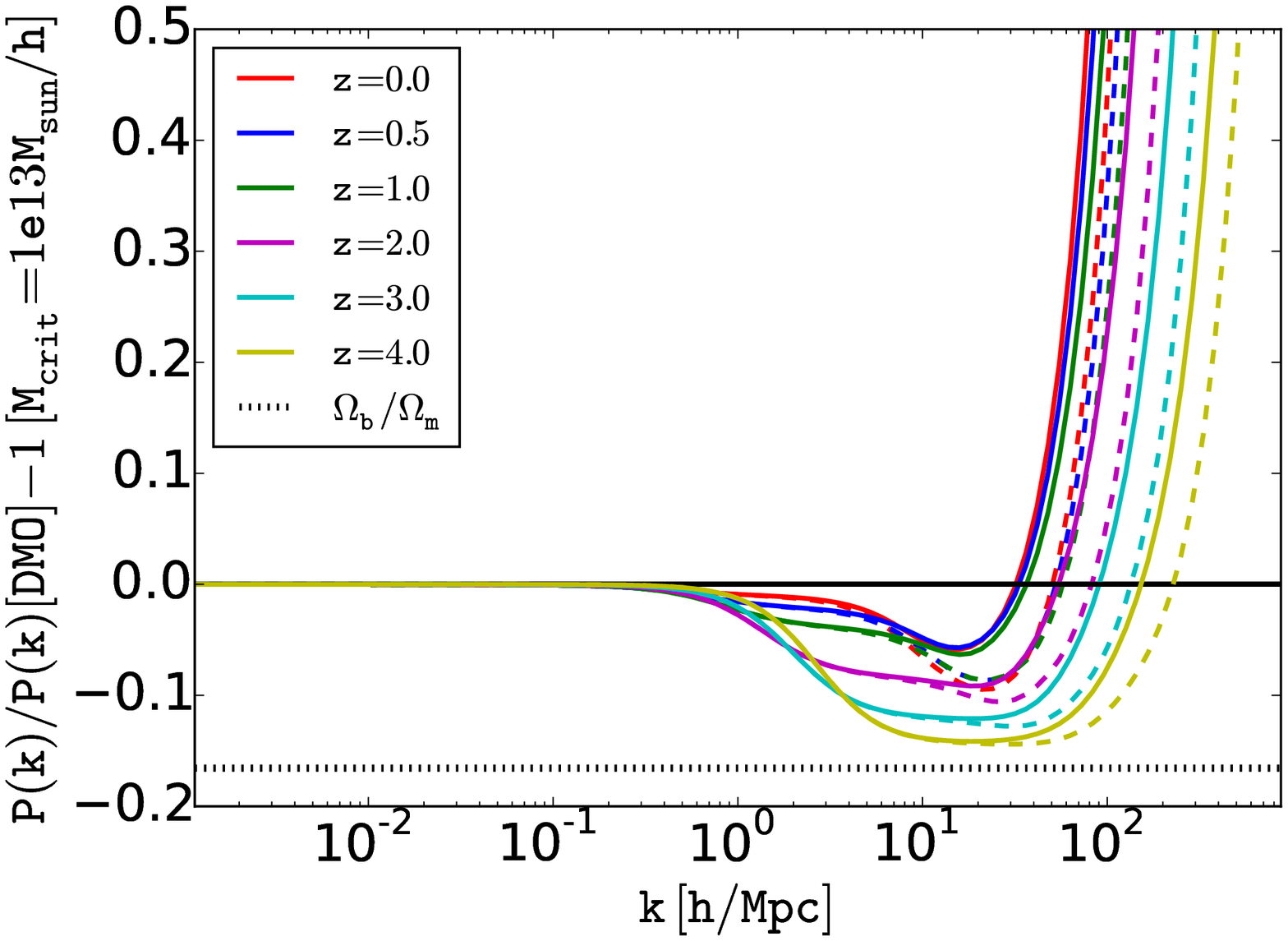}
	\includegraphics[width=0.48\textwidth]{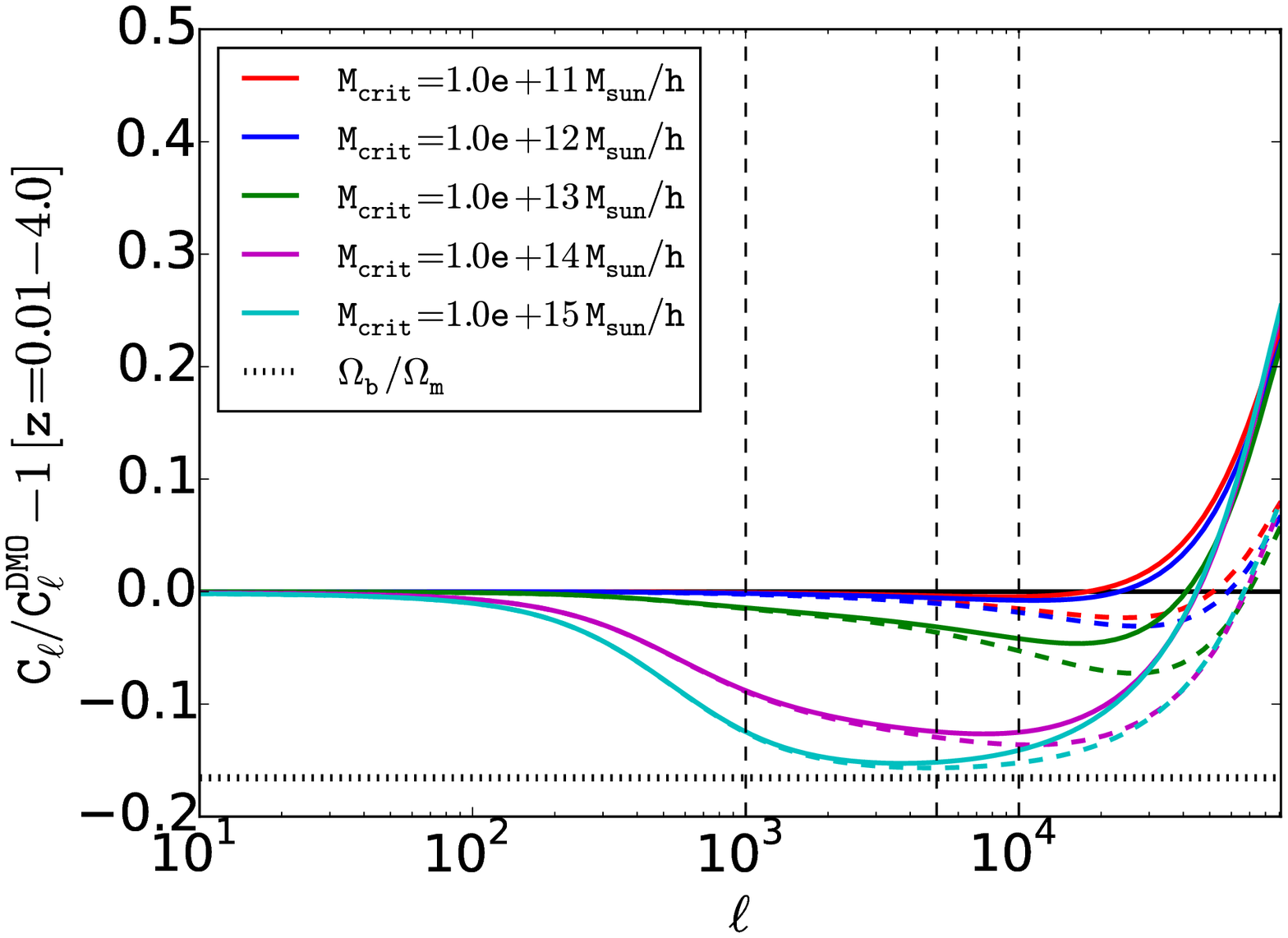}
	\includegraphics[width=0.48\textwidth]{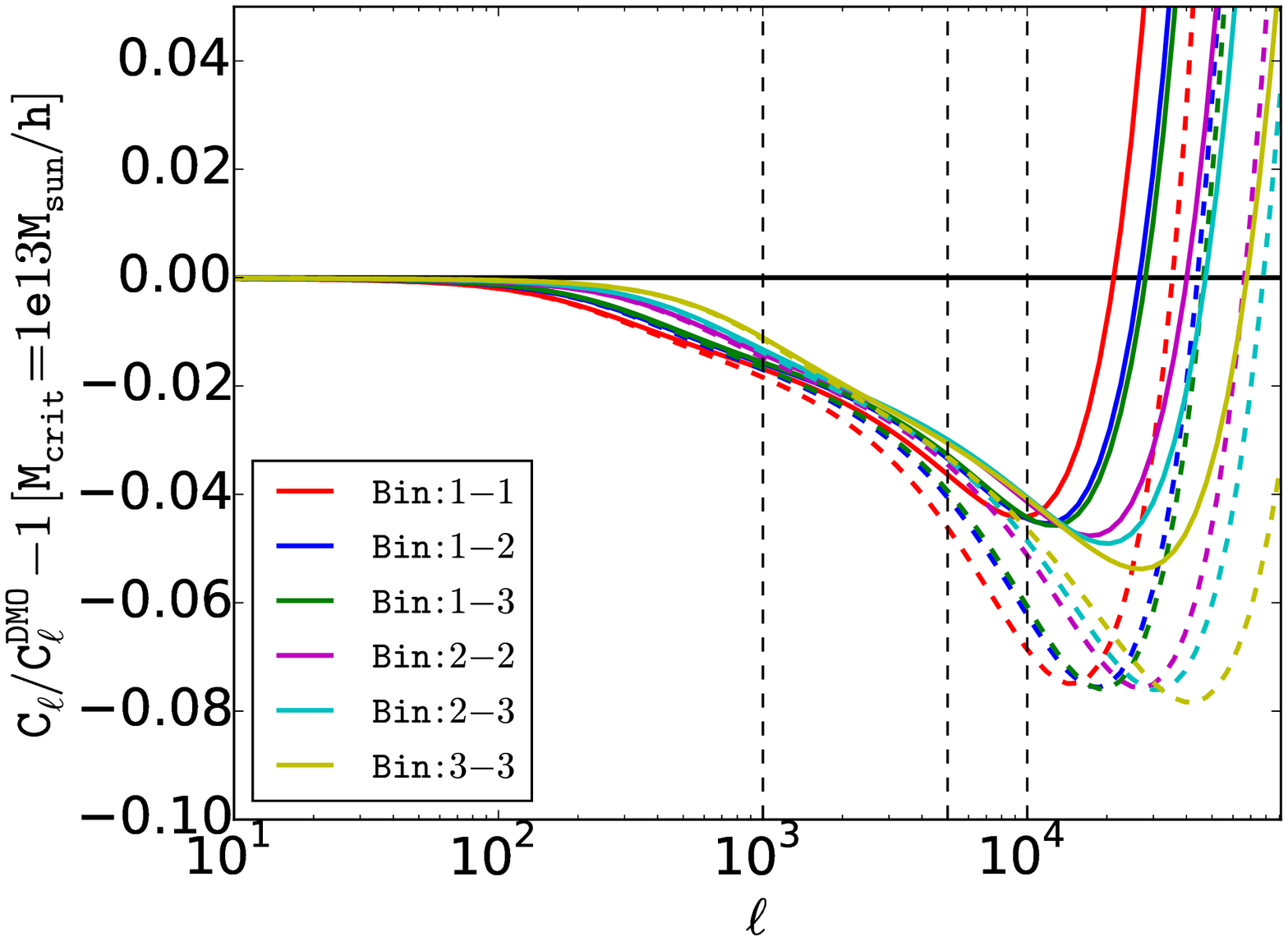}

	\caption{Top row: Relative deviation of the matter power spectrum predicted by the BAR model from the DMO model predictions as a function of $k$ for different $M_{\rm crit}$ at redshift zero (left) and for fixed $M_{\rm crit} = 10^{13} h^{-1} M_{\bigodot}$ and different redshifts (right). Bottom row: Relative deviation of the shear power spectrum ($C_{\ell}$) predicted by the BAR model from DMO model predictions for different $M_{\rm crit}$ in one big redshift bin (left) and for three tomographic redshift bins and fixed $M_{\rm crit} = 10^{13} h^{-1} M_{\bigodot}$ (right). Dashed lines are the calculations without adiabatic contraction (AC) and solid lines with adiabatic contraction (AC). The horizontal dashed line shows the cosmic baryon fraction.}
	\label{fig:pkandcl}
\end{figure*}

In this section we try to draw a comparison between the baryonic model (BAR) and the dark-matter only (DMO) model. We would like to establish an understanding of the scales where the baryonic corrections become important and how these scales changes with redshift and the only free parameter, $M_{\rm crit}$.

Figure \ref{fig:pkandcl} (top-left panel) shows the relative differences between the BAR and DMO predictions for the matter power spectrum, also referred as {\it boost} in this article. There is only one free parameter of the baryonic model, $M_{\rm crit}$ which regulates the amount of AGN feedback and which is introduced in section \ref{sec:icm}. The overall shape of the deviation is similar in all cases for various $M_{\rm crit}$ and redshifts: the BAR model follows the DMO model for large scales, suffers a deficit in power at intermediate scales due to flatter gas profile compared to the NFW profile and finally the power shoots up due to the central stellar component. Also without adiabatic contraction (AC) the raise in the matter power spectrum occurs at very small scales, but including AC effect this raise can be seen at comparatively lower $k$ or larger scales. This is because AC makes the profile steeper in the centre and shallower in the outskirts.

At redshift 0 (top-left panel of figure \ref{fig:pkandcl}), the baryonic correction starts showing up (more than 1\%) at $k \sim 5\  h/Mpc$ for models with negligible AGN feedback (lower $M_{\rm crit}$), whereas for more extreme AGN feedback models (higher $M_{\rm crit}$) this correction is important at much larger scales like $k \sim 0.1\  h/Mpc$. In our fiducial BAR model with $M_{\rm crit} = 10^{13} h^{-1} M_{\bigodot}$, the baryonic effects become significant, i.e., more than 1 percent, at $k \sim 0.5\  h/Mpc$. The maximum dip in the intermediate scales vary for different $M_{\rm crit}$; for the most extreme models where AGN feedback can push all the gas out of the halo, this dip is nearly the cosmic baryon fraction, $\Omega_b/\Omega_m$. However, for more a realistic model ($M_{\rm crit} = 10^{13} h^{-1} M_{\bigodot}$) this dip is nearly 7-8\%. For more optimistic models like $M_{\rm crit} = 10^{12} h^{-1} M_{\bigodot}$, this dip is even smaller, nearly 4-5\%. Therefore, we can conclude the more extreme AGN feedback models triggers the deviation of matter power spectrum from DMO model at larger scales and also the dip in the power at intermediate scales can be as large as the cosmic baryon fraction in case where all the gas are pulled out by the AGN feedback, however, for more realistic and optimistic models, the deviation starts at relatively small scales and also the maximum dip is comparatively smaller.

Figure \ref{fig:pkandcl} (top-right panel) shows the same quantity for a fixed $M_{\rm crit} = 10^{13} h^{-1} M_{\bigodot}$ at different redshifts. If we go to higher redshift, the overall shape of the deviation of the BAR matter power spectrum from the prediction of the DMO model (boost) is nearly the same as at redshift zero, however, the scales and the maximum dip amplitude at various redshifts change. We see that at higher redshifts, the dip starts to trigger at larger scales and also the maximum dip converge to the cosmic baryon fraction.

In figure \ref{fig:pkandcl} (bottom-left panel), the baryonic correction to $C_{\ell}$ is shown in one big redshift bin ($z=0.01-4.0$). Here, the shear power spectrum starts to deviate from DMO predictions at about $\ell =100$ for the most extreme AGN feedback models and at $\ell$ of about several thousands for models with weak AGN feedback. For our realistic model (green curve), this deviation occurs at about $\ell \sim 700$. The maximum dip in power is very similar to that of the matter power spectrum explained above. It is worth noticing that for $\ell=10000$ the deviation is very significant for the realistic model ($M_{\rm crit} = 10^{13} h^{-1} M_{\bigodot}$), however, it is negligible for the optimistic model ($M_{\rm crit} = 10^{12} h^{-1} M_{\bigodot}$). Because these are the cases that we study in our likelihood analysis, we will show in section \ref{sec:cosmology} that this behaviour is consistent with the cosmological parameter estimation with these models.

%===============================================================================

%===============================================================================
\section{Fiducial model and mock datasets}
\label{sec:fiducial}

\begin{figure*}
  \includegraphics[width=0.48\textwidth]{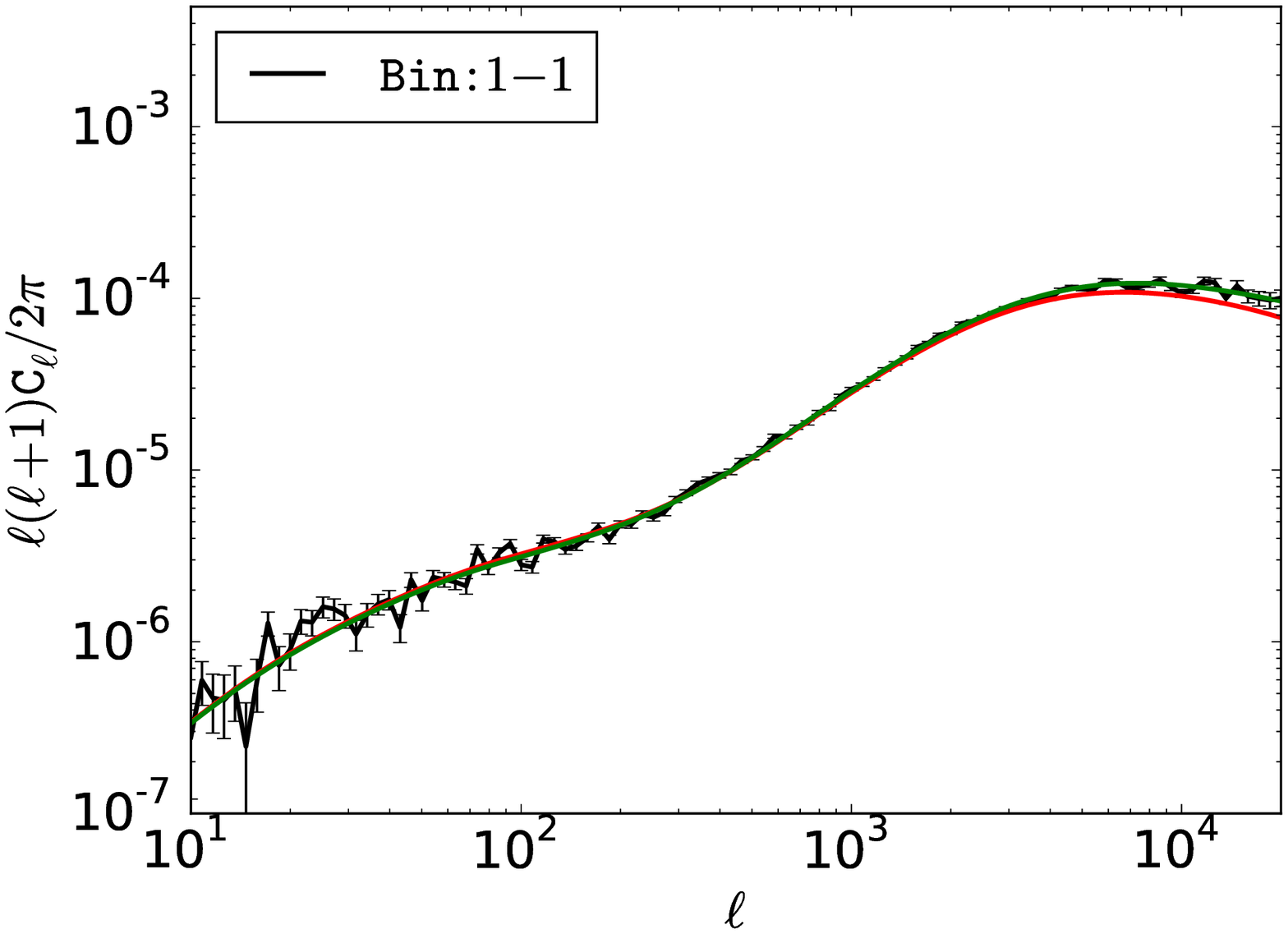}
  \includegraphics[width=0.48\textwidth]{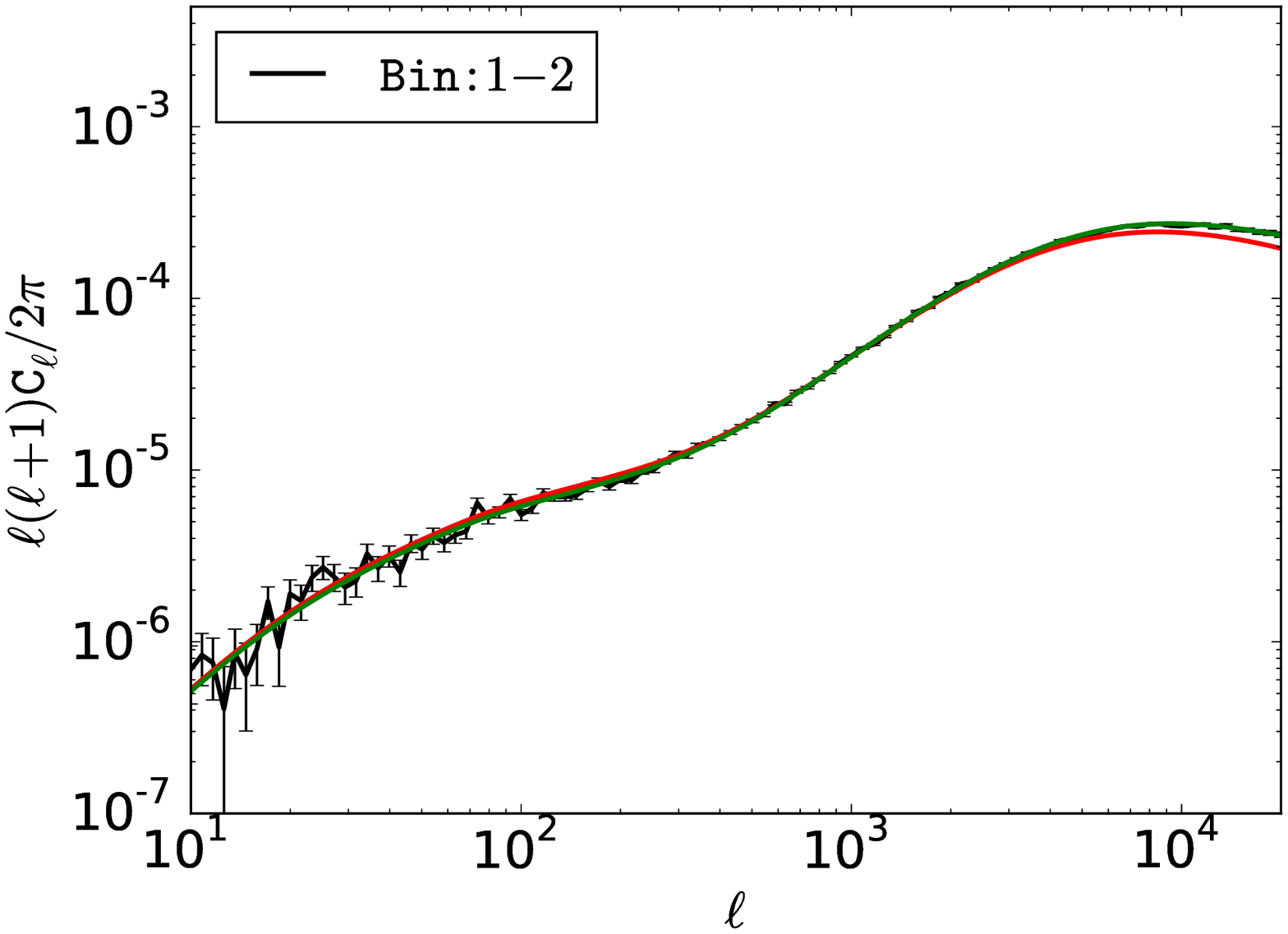}
  \includegraphics[width=0.48\textwidth]{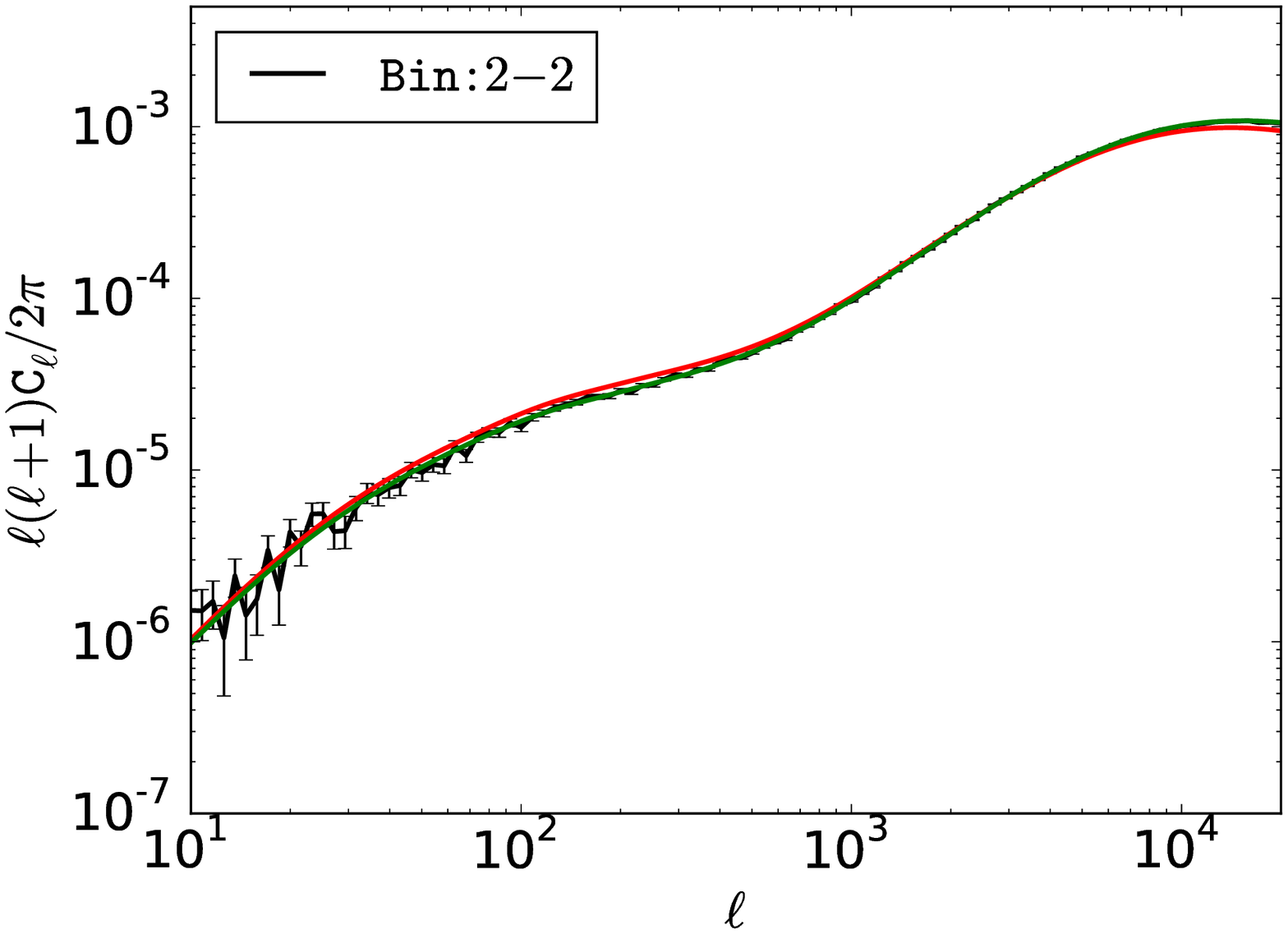}
  \includegraphics[width=0.48\textwidth]{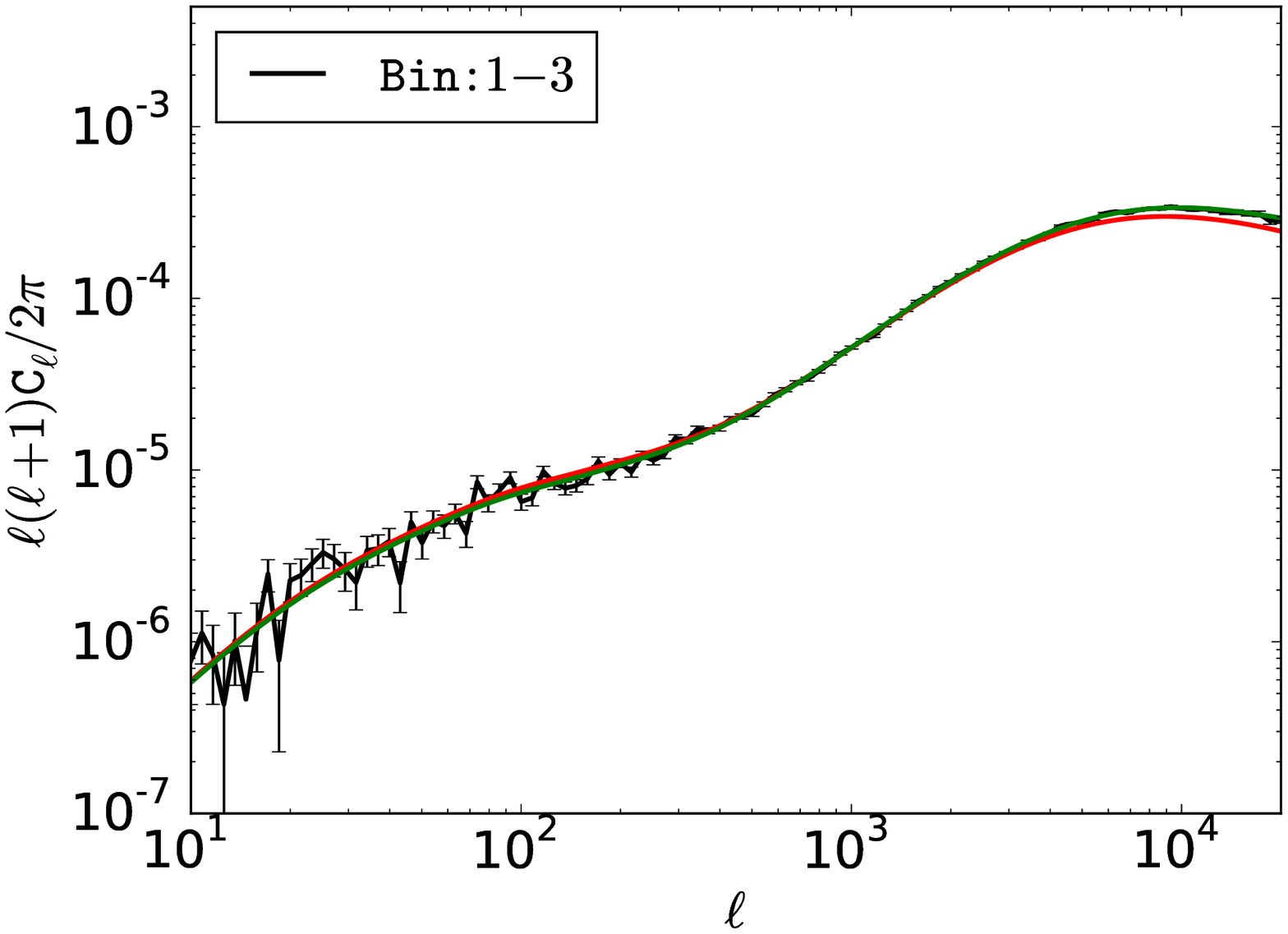}
  \includegraphics[width=0.48\textwidth]{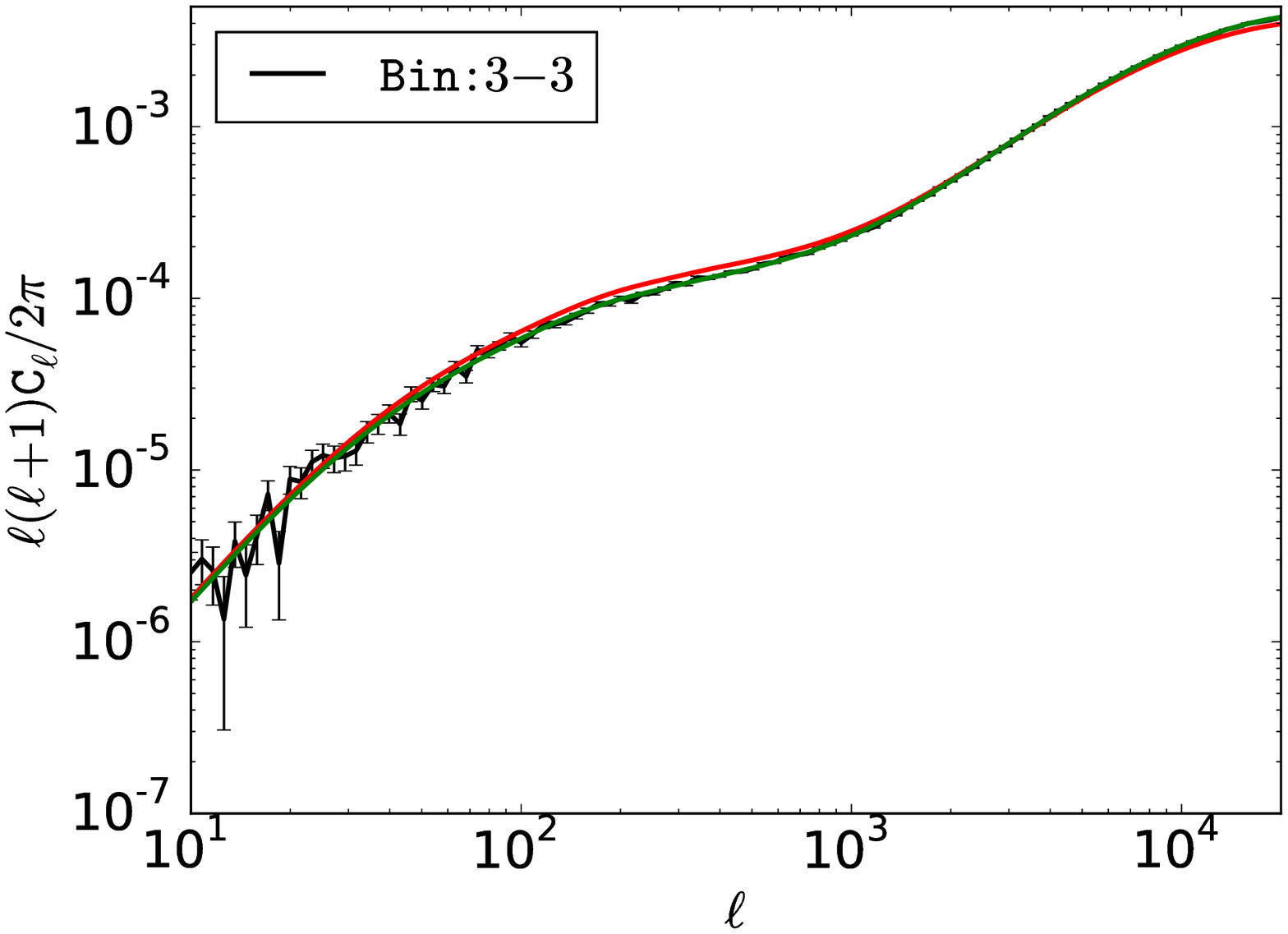}
  \includegraphics[width=0.48\textwidth]{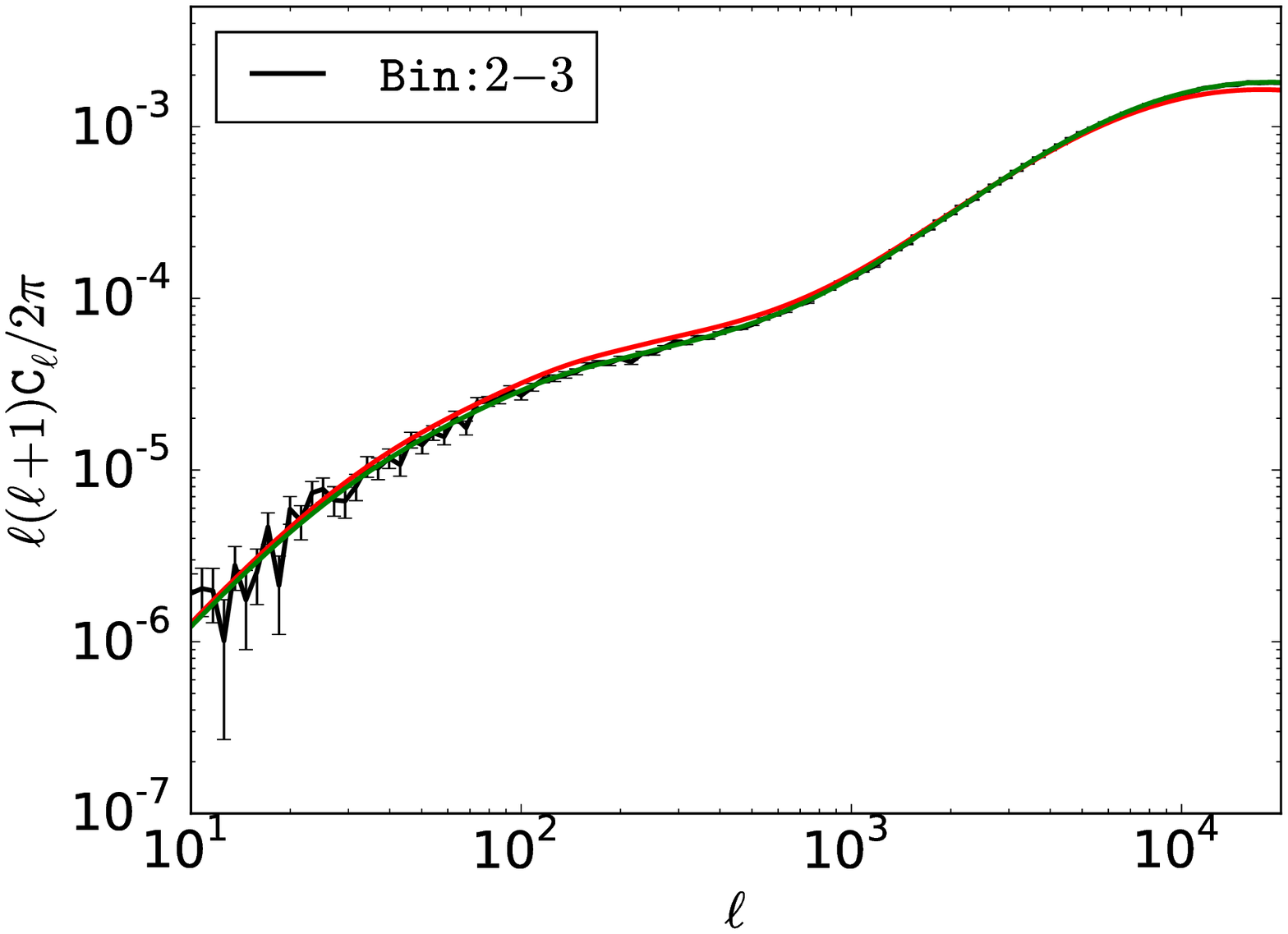}

  \caption{Mock datasets (including random noise) for $\ell_{max}=20000$ in all six spectra (in black). The left column shows the three auto-spectra and the right column shows the three cross-spectra. Solid lines show the best fit for the DMO (red) and BAR (green) models.}
  \label{fig:bestfit}
\end{figure*}

In this section, we would like to mention two factors that are quite important for our experiments - fiducial parameters and mock datasets.  The fiducial parameters assumed in this work, particularly about cosmology, baryonic model and Euclid mission, are very standard. Also the mock datasets generated are correctly contaminated with random noise. Following are the key numbers and information about the fiducial model assumed and mock datasets:

\begin{enumerate}

\item We used WMAP - 5th year cosmology  as our fiducial model with $[\Omega_m, \Omega_b, h, n_s, \sigma_8, w_0, w_a]$ as $[0.279, 0.0462, 0.701, 0.96, 0.817, -1.0, 0.0]$. We assume the equation of state of dark-energy is redshift dependent as \citep{2001IJMPD..10..213C,2003PhRvL..90i1301L},
\begin{equation}
	w(a) = w_0 + (1-a)w_a
\end{equation}
\\
where, $a = 1/(1+z)$ is the scale factor at redshift $z$.

\item We used three redshift bins to do the tomographic analysis with boundaries $[0.01, 0.8, 1.5, 4.0]$. So we calculated a total of six spectra - three auto-spectra between bins 1-1, 2-2 and 3-3 and three cross-spectra between bins 1-2, 1-3 and 2-3. 

\item We perform the likelihood analysis for different $\ell_{max}$ with $\ell_{min}=10$ and 100 equally spaced logarithmic bins. So the bin sizes for the likelihood analysis with different $\ell_{max}$ are different.

\item We assumed that the mean redshift of the source distribution to be nearly 1.0 which gives approximately 50 galaxies per arc min$^2$ and $f_{\rm sky}=0.55$ which resembles Euclid like survey.

\item For the baryonic model, we used the realistic AGN feedback model $M_{\rm crit}= 10^{13} h^{-1}M_{\bigodot}$ as the fiducial value for total nine $\ell_{max}$ (1000, 2000, 3000, 4000, 5000, 6000, 8000, 10000, 20000). We also performed one case with more optimistic model $M_{\rm crit} = 10^{12} h^{-1}M_{\bigodot}$ for $\ell_{max}=10000$. So there are ten cases in total. 

\item We used our fiducial model stated above to generate shear power spectrum $C_{\ell}$ for these ten cases and perturbed all $C_{\ell}$ with normally distributed multi-variate random numbers drawn from a distribution with mean $C_{\ell}$ and the corresponding covariance matrix. These $C_{\ell}$ are catalogued and constitute the mock data sets. So, there are total ten mock data sets. In figure \ref{fig:bestfit} we show the mock datasets up to $\ell_{max}=20000$ for the six spectra and the best fits (which will be discussed in section \ref{subsec:goodness}).

\end{enumerate}

For each bin combination (1-1,1-2 etc), the length of the data vector ($\ell$ or $C_{\ell}$) is 100. Therefore, the total number of data points in each data set is 600. However, the two cross-spectra, 1-2 and 1-3, are highly correlated which actually leads us to have only 5 degree of freedom for each $\ell$. Therefore, the total number of degree of freedom in each data set is about 492 (500 - 8 free parameters). Hence, the best fit to each dataset can have a $\chi^2$ in the range 492 $\pm \sqrt{(2\times 492)}$ which is between 470 and 514.

In figure \ref{fig:pkandcl} (bottom-right panel), we show the boost for the unperturbed (without random noise) mock datasets up to very high $\ell_{max}$ with the corresponding DMO model. In all six curves of this figure, we kept $M_{\rm crit}=10^{13} M_{\bigodot}$. The auto-spectra in the first bin (1,1), starts deviating (more than 1\%) from the DMO model at about $\ell=300$ whereas the auto-spectra of the third bin (3,3) starts showing deviation at nearly $\ell=800$. All other auto-spectra and cross-spectra are between these two extremes. This behaviour is justified by looking at the same figure in upper-right panel, which shows the redshift evolution of the correction for the same $M_{\rm crit}$. It can be seen that at higher redshifts, the BAR matter power spectrum starts to deviate from DMO at smaller scales but also induces a larger dip at intermediate scales due to gas expulsion. This behaviour can be seen in the bottom-right panel. The $C_{\ell}$ in the lower redshift bin (1-1) starts deviating from DMO at larger scales as compared to the higher redshift bin (3-3), but the maximum dip in the two cases can be seen in the higher redshift bin (3-3). If we compare this to the bottom-left panel of the same figure, one can notice that the baryonic correction becomes even important when binning in redshift rather than using one big redshift bin. This provides additional constraints on $M_{\rm crit}$ while performing the analysis in tomographic bins compared to poorer constraints when only one bin is used.

%===============================================================================
\begin{figure*}
\centering

  \includegraphics[width=0.8\textwidth]{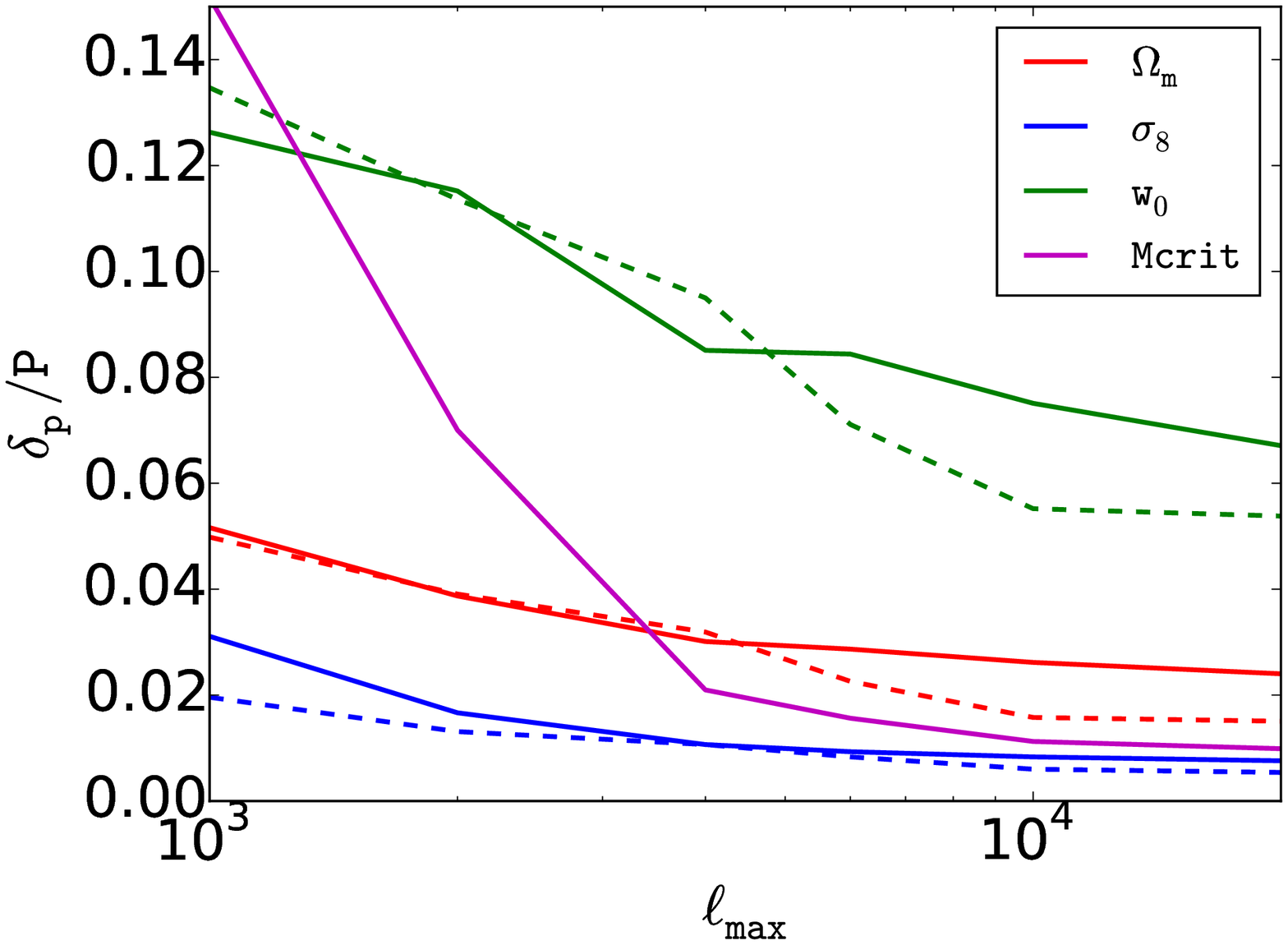}
  \caption{Relative 1$\sigma$ errors on different cosmological parameters as a function of $\ell_{max}$ for $M_{\rm crit}=10^{13} h^{-1} M_{\bigodot}$. Solid lines are for the BAR model and dashed curves are for the DMO model. Horizontal black dashed lines mark the $\pm 1$ and vertical black dashed lines shows important scales.}
  \label{fig:errors}
\end{figure*}

\section{Likelihood analysis and cosmological implications}
\label{sec:cosmology}

We performed a likelihood analysis using MCMC to explore the cosmological parameter space for nine different $\ell_{max}$ (1000, 2000, 3000, 4000, 5000, 6000, 8000, 10000, 20000) using $M_{\rm crit}=10^{13} h^{-1} M_{\bigodot}$, which is our most realistic model, and for $\ell_{max}=10000$ using $M_{\rm crit}=10^{12} h^{-1} M_{\bigodot}$ which is our optimistic model. 

We run MCMC on the ten mock datasets obtained adopting both the DMO and BAR models, therefore we run a total of 20 MCMC. Each MCMC is performed using the publicly available code COSMOMC \citep{2002PhRvD..66j3511L}, with 16 chains in each case. So total 320 CPUs are used for nearly 10 days to reach the desired convergence. The whole analysis required about 76800 hours.

We demonstrate the results of the MCMC and the interpretation in the following two sections, targeting particularly the precision and accuracy in predicting the cosmological parameters. 

\begin{figure*}
\centering
  \includegraphics[width=0.48\textwidth]{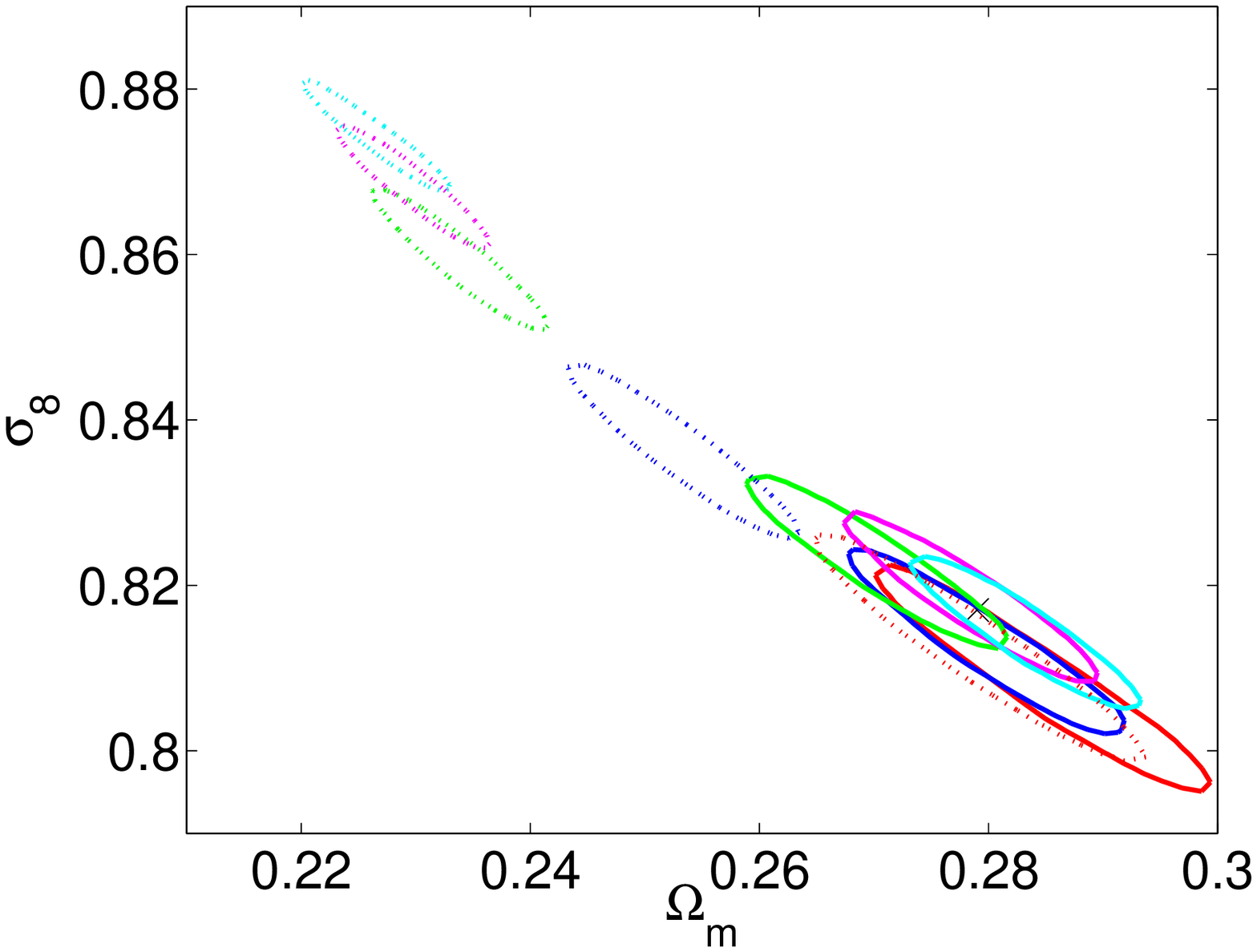}
  \includegraphics[width=0.48\textwidth]{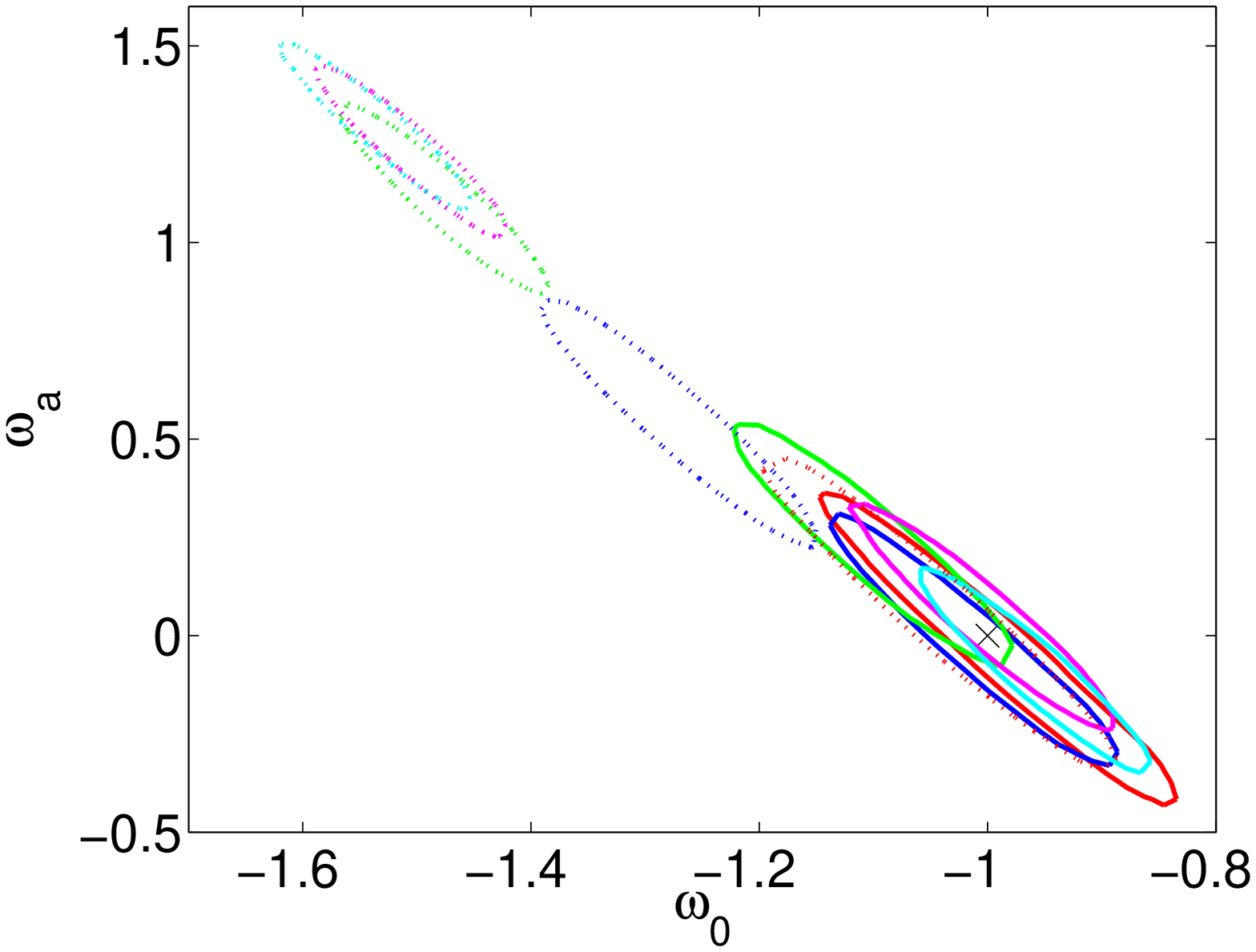}\\
  \includegraphics[width=0.48\textwidth]{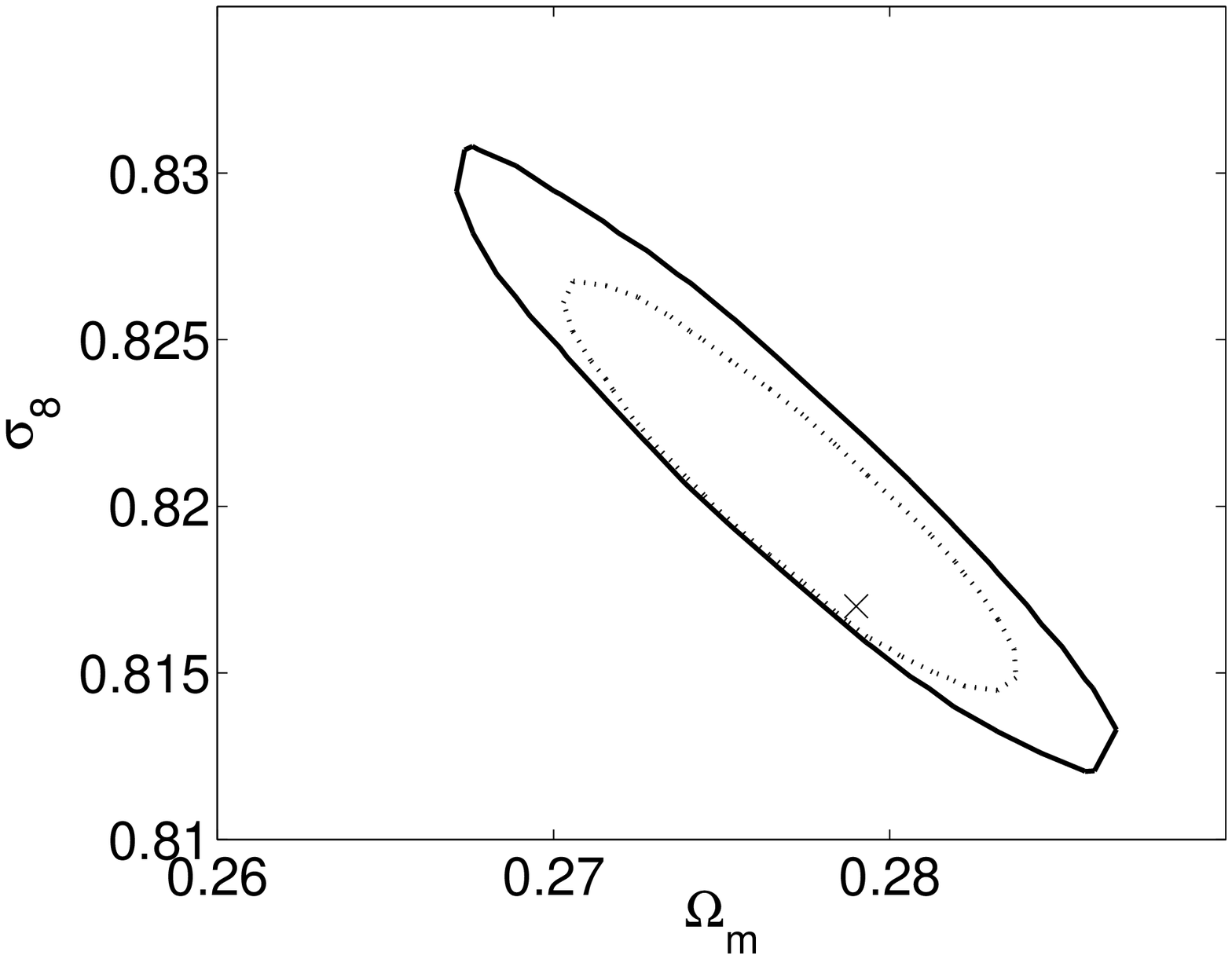}
  \includegraphics[width=0.48\textwidth]{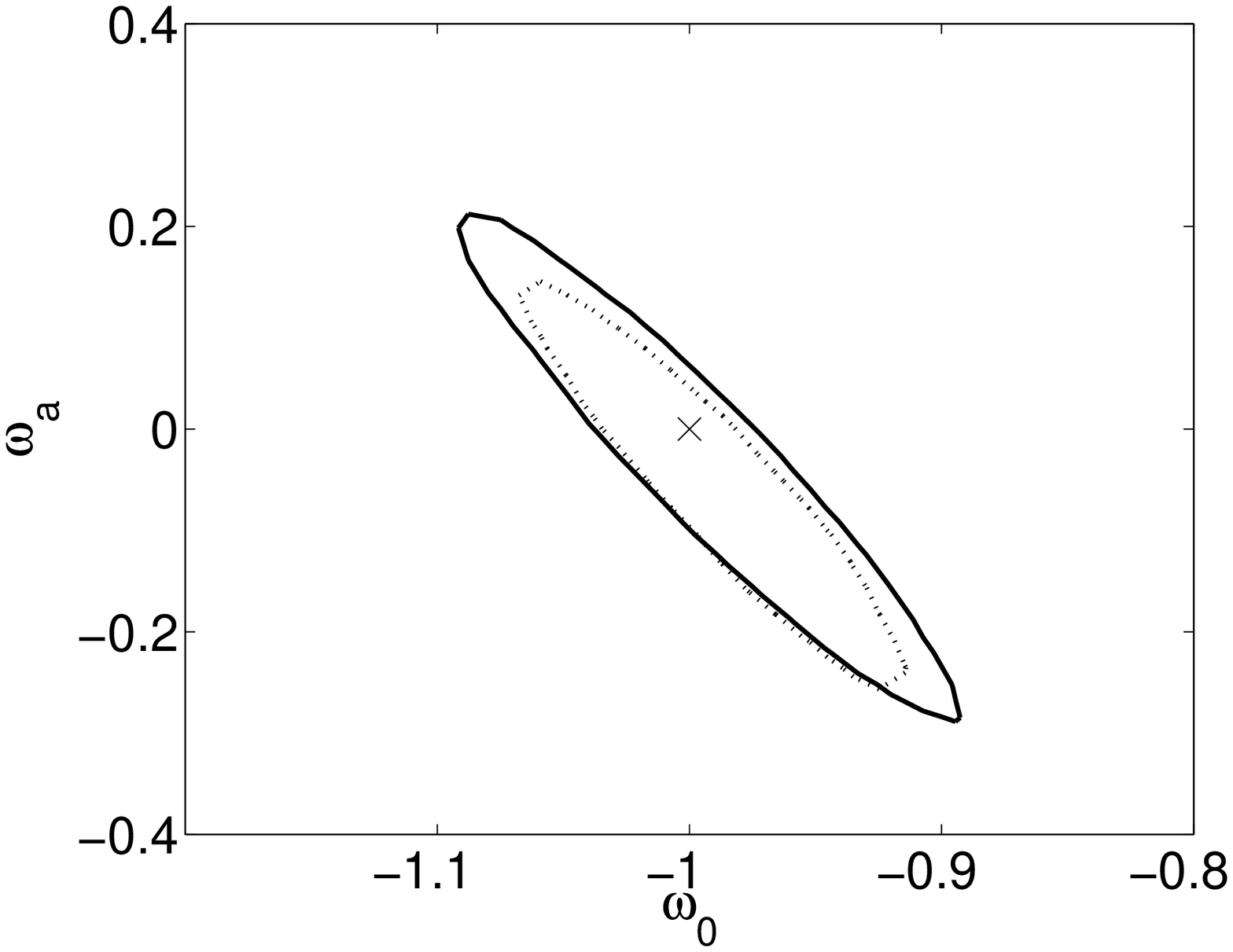}

  \caption{Top row: 1$\sigma$ 2D error ellipses for different cosmological parameters using mock datasets with $M_{\rm crit}=10^{13} h^{-1} M_{\bigodot}$ and different $\ell_{max}=$ 3000 (red), 5000 (blue), 8000 (green), 10000 (magenta), 20000 (cyan). Bottom row: 1$\sigma$ 2D error ellipses using mock datasets with $M_{\rm crit}=10^{12} h^{-1} M_{\bigodot}$ for $\ell_{max}=$ 10000. All solid curves are for the BAR model and dashed curves are for the DMO model.}
  \label{fig:mcmc}
\end{figure*}

%---------------------------------------------------------------------------------------------

\subsection{Precision in cosmology}

Future experiments, like Euclid, are expected to provide very tight constraints on cosmological parameters. Here we show the constraints expected from using the weak lensing shear power spectrum as a function $\ell_{max}$. Figure \ref{fig:errors} shows the relative variance of four cosmological parameters and one baryonic parameter using both models, BAR (solid curves) and DMO (dashed curves). The matter density of the Universe ($\Omega_m$) and the amplitude of fluctuations ($\sigma_8$) are the most constrained parameters, however, other parameters like the equation-of-state of dark-energy today ($w_0$) are relatively less constrained. The overall behaviour of all parameters is the same, weak constraints for small $\ell_{max}$, better constraints with increasing $\ell_{max}$ and a flattening beyond $\ell_{max}\sim 8000$. The constraints derived from the BAR model are relatively weaker than the constraints derived from DMO model, which is the consequence of the extra parameter, $M_{\rm crit}$. 

The normalized matter density of the Universe $\Omega_m$ can already be determined up to 5\% at $\ell_{max}$=1000 which improves as good as 2-3\% at $\ell_{max}$=8000 whereas the amplitude of fluctuations $\sigma_8$ can be determined much better at corresponding scales. At $\ell_{max}$=1000, $\sigma_8$ can be known up to 3\% and these constraints improves better than 1\% at $\ell_{max}$=8000. After $\ell_{max}$=8000, the variance of both the parameters remains the same and no further constraints can be drawn by going up to lower scales or higher $\ell_{max}$. There is a certain degeneracy in these two parameters which can be seen in figure \ref{fig:mcmc} upper-left panel, where different colours represent different $\ell_{max}$.

The constraints on the two parameters describing the redshift evolution of the equation of state of dark energy, $w_0, w_a$ can also be improved with this kind of experiments. At $\ell_{max}$=1000 $w_0$ can only be determined as good as 12\%, whereas for $\ell_{max}$=8000 it can be constrained up to 6-7\% and with the same precision for higher $\ell_{max}$. However, the constraints on $w_a$ are much weaker. The absolute error on $w_0$ is nearly 0.35 for $\ell_{max}$=1000, $\sim$0.18 for $\ell_{max}$ = 8000 and the same afterwards. 

The flattening of the relative errors of the parameters indicates that there is no gain in precision of cosmological parameters estimation after a certain threshold $\ell_{max}\sim 8000$. In practice, an experiment like Euclid may provide us with very high quality data to even resolve and measure the shear power spectrum at $\ell_{max}$ as high as $10^5$, but our analysis shows that the constraints becomes constant after $\ell_{max}\sim 8000$ and no further improvement can be achieved.

This forecast suggests that by measuring $C_{\ell}$s up to $\ell_{max}\sim 8000$, one can constrain $\Omega_m$ to about 2\% precision and $\sigma_8$ to about 0.5\% precision without any loss of information from high $\ell$s and including baryonic physics. However, $w_0$ can only be constraints up to 6-7\% with some information about $w_a$, the time derivative of the equation of state of dark energy.

\begin{figure*}
\centering

  \includegraphics[width=0.8\textwidth]{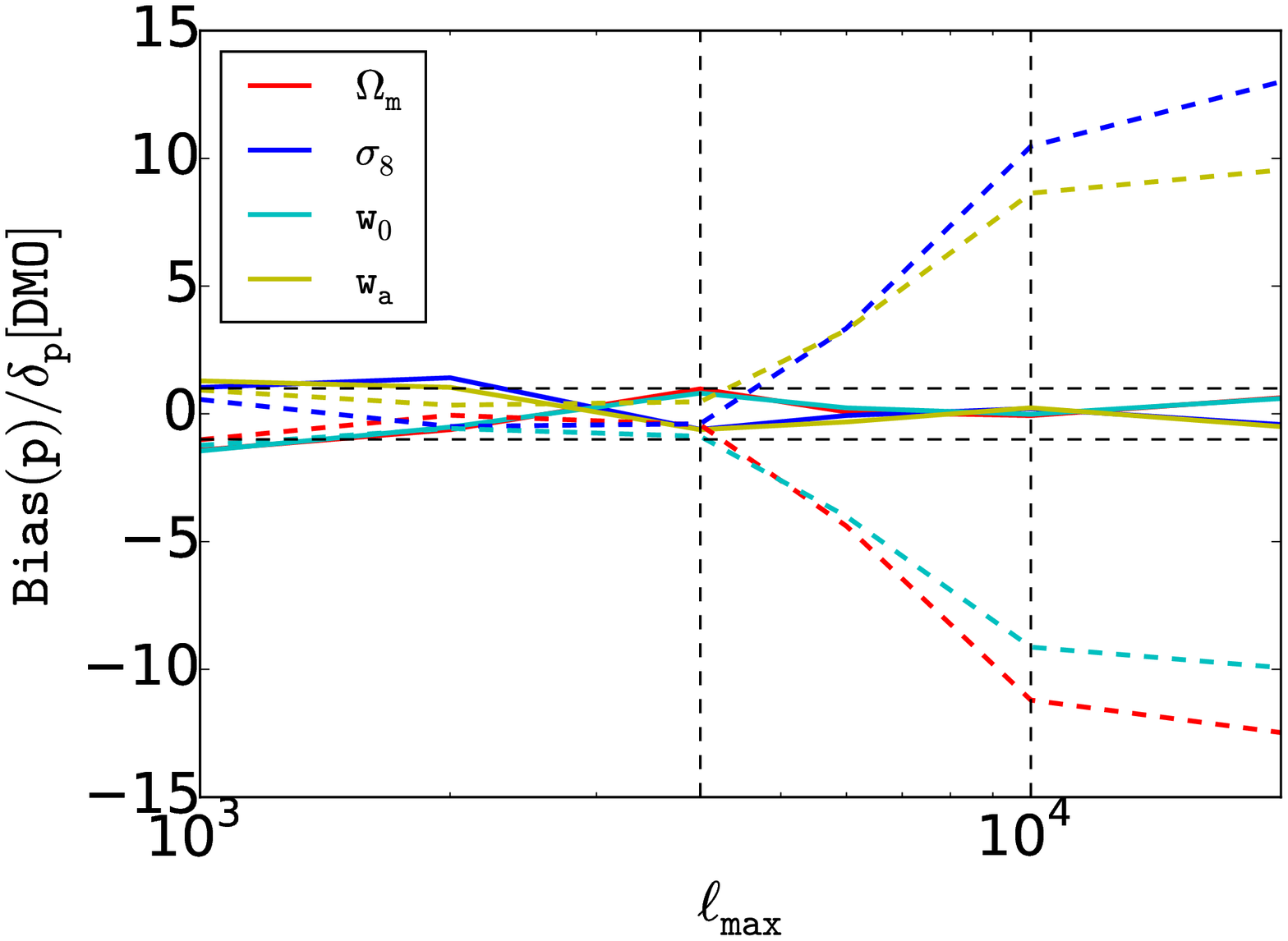}
  \caption{The ratio of bias and 1$\sigma$ error of various cosmological parameters as a function of $\ell_{max}$ for $M_{\rm crit}=10^{13} h^{-1} M_{\bigodot}$. Solid lines are for the BAR model and dashed curves are for the DMO model. Horizontal black dashed lines mark the $\pm 1$ and vertical black dashed lines shows important scales.}
  \label{fig:bias}
\end{figure*}

%---------------------------------------------------------------------------------------------

\subsection{Accuracy in cosmology}

When precision cosmology is the goal, one should also take into account the ability to recover the cosmological parameter accurately. If there are systematic errors in the model, one can still derive very tight constraints from the wrong model, but the recovered parameters will be wrong or biased as compared to the true values. In this section we will present the results from our analysis of the bias in the cosmological parameters due to the lack of baryonic physics in DMO models and we will assess if these biases are significant. We define bias as the difference between the mean value of the parameter in MCMC and its fiducial or true value.

Figure \ref{fig:mcmc} shows the 1$\sigma$ error ellipses of cosmological parameters when the model is BAR (solid curves) and DMO (dashed curves). For small $\ell_{max}$, the two models are indistinguishable, a consequence of the fact that baryonic physics becomes more important only at smaller scales. But as we go higher and higher in $\ell_{max}$, the target density of the DMO model shifts further from the true target density, however, the BAR model remains at the correct location.  We find that for all BAR models this bias is smaller than the 1$\sigma$ error of the parameter, however, the bias in the parameters obtained fitting for the DMO model increases with increasing $\ell_{max}$.

Figure \ref{fig:bias} shows the ratio of these biases and the 1$\sigma$ error on the cosmological parameters as a function of $\ell_{max}$ for the two models, BAR (solid curves) and DMO (dashed curves). The bias never exceeds the 1$\sigma$ error for the BAR models, however, it does for the DMO models only after $\ell_{max} \sim 4000$. This is again a consequence of the fact that baryonic physics is only important at smaller scales. This indicates that if we only perform our experiment up to $\ell_{max}$=4000, no baryonic physics needs to be taken into account, however, if one is interested in $\ell_{max}>4000$ baryonic physics becomes very important. After $\ell_{max}$=4000 the bias increases with $\ell_{max}$ and goes as big as 10$\sigma$ at $\ell_{max}$=10000 and remain flat after that. We see in the previous section that constraints on cosmological parameter can still be improved up to $\ell_{max}$=8000, but considering the wrong model, DMO, the cosmological parameters will be 5-10$\sigma$ away from the true values. So, in order to gain the best constraints on cosmology, baryonic physics must be taken into account.

%---------------------------------------------------------------------------------------------

\subsection{An optimistic model}
We analysed the $\ell_{max}=10000$ case for our optimistic model with $M_{\rm crit}=10^{12} h^{-1} M_{\bigodot}$. As in our previous analysis, we performed two MCMC in this case too, fitting for the BAR model and for the DMO model. Figure \ref{fig:mcmc} (bottom row) shows the 1$\sigma$ error ellipses of cosmological parameters. In this case the bias in the cosmological parameters does not exceed the $1\sigma$ error and hence is not a very troubling case. This was expected, as for lower $M_{\rm crit}$, baryonic physics is less important even at comparatively small scales as compared to cases where $M_{\rm crit}$ is higher. For example if we look at figure \ref{fig:pkandcl} (bottom-left panel), we can see that for $M_{\rm crit}=10^{12} h^{-1} M_{\bigodot}$, the deviation of $C_{\ell}$ from the DMO model is negligible at $\ell=10000$. Hence, we actually expect smaller or no bias. 

%---------------------------------------------------------------------------------------------

%===============================================================================

\section{Discussion and conclusions}
\label{sec:discussion}

In this work we first review the important theoretical framework necessary to calculate the matter power spectrum using the halo model and to compute the shear angular power spectrum in different redshift bins. We presented an analytic prescription to distribute baryons into two components -- the intra-cluster plasma in hydrostatic equilibrium within the halo, and the BCG, which dominates the mass distribution in the centre of the halo, and whose properties are well measured using abundance matching techniques. We also take into account the adiabatic contraction of the dark matter particles due to the central condensation of  baryons. We also compared these analytic density profiles to the simulations of \cite{2014MNRAS.440.2290M}, both dark-matter-only and baryonic with AGN feedback, and found a remarkable agreement. 

We model the shear power spectrum in the two models, BAR and DMO, and found that baryonic corrections are important after $k \sim 0.5\ h/Mpc$ in the matter power spectrum at redshift 0 for our most realistic AGN feedback model, which translates into $\ell \sim 800$ for the shear power spectrum in one big redshift bin. However, if binned in redshift space (lensing tomography), these corrections become larger in each bin and for each auto- and cross-correlation function. These baryonic corrections have one free parameter, $M_{\rm crit}$, which regulates AGN feedback, i.e., it controls how much gas will be inside the halo as a function of the halo mass. We believe the most realistic value of this parameter is near $10^{13} h^{-1}M_{\bigodot}$, which sets the most likely magnitude of  baryonic corrections.

We perform the likelihood analysis using MCMC for total ten different datasets. Nine of them assume our realistic model for the AGN feedback with $M_{\rm crit} = 10^{13} h^{-1}M_{\bigodot}$ but different $\ell_{max}$, and one assumes a less extreme (optimistic) model with $M_{\rm crit} = 10^{12} h^{-1}M_{\bigodot}$. For each mock dataset, we perform MCMC to fit for both models, BAR and DMO.

The main results of the likelihood analysis are summarized in figure \ref{fig:mcmc}, \ref{fig:errors} and \ref{fig:bias}. The results are very interesting in two aspects: first, we found that the constraints on all cosmological parameters improve with increasing $\ell_{max}$, but after $\ell_{max} \sim 8000$, the variance of each parameter becomes nearly constant. This indicates that even if we go to higher $\ell_{max}$ (or smaller scales), no additional constraints on the cosmological parameters can be gained. Second, if the wrong model, in this case DMO, is fitted to the data, after $\ell_{max}=4000$ the mean recovered value of the parameters starts moving away from its true value. We refer to the difference between the true value and recovered mean value as bias in the cosmological parameter. The bias in the parameters becomes more than 1$\sigma$ after $\ell_{max}=5000$ and goes up to 10$\sigma$ for $\ell_{max}=10000$, remaining flat afterwards. So, there is a very interesting window from $\ell_{max}=4000-8000$ which is useful for improving the constraints on cosmology, but if wrong model like DMO is chosen, the recovered cosmology can be highly biased from few to 10-$\sigma$.

\subsection{Goodness of fit}

\begin{figure}
  \includegraphics[width=0.48\textwidth]{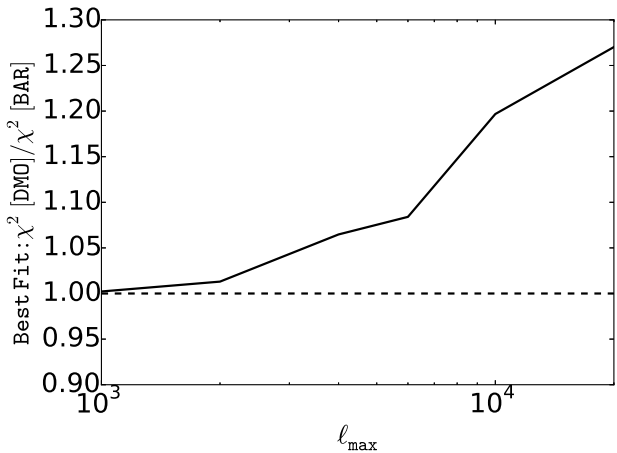}
  \caption{Showing the ratio of the best fit $\chi^2$ in DMO model and BAR model at different $\ell_{max}$.}
  \label{fig:chi2}
\end{figure}

In the previous sections we see that for $\ell_{max}<4000$, there is no significant bias added to the determination of the cosmological parameters in our analysis, however, for $\ell_{max}>5000$ the bias exceeds 1$\sigma$ and keep increasing up to 10$\sigma$ with increasing $\ell_{max}$. The question here is: can we discard these biased models by looking at the goodness of fit? The answer to this question lies in figure \ref{fig:chi2} where we show the ratio between the best fit $\chi^2$ in the DMO model and that in the BAR case as a function of $\ell_{max}$. This ratio is as little as 5-10\% up to $\ell_{max} \sim 5000$ but after that it only goes up to 25\% at $\ell_{max}=20000$ where bias is more than 10$\sigma$. Now, the reduced $\chi^2=1.25$ does not appear as such a bad fit for our cosmological measurements. So, by looking at the $\chi^2$ only, it is not really possible to discard a model. The same conclusion can be drawn from figure \ref{fig:bestfit}, where we show the mock datasets of the six spectra (between different bins) for $\ell_{max}=20000$. In this figure, we also show the two best fit from the DMO model (in red) and the BAR model (green). As we expect, the green curve is a better fit to the data than the red curve. But if the green curve is not present in this figure, the red curve does not appear to be a very bad fit. So, when deriving constraints on cosmology from this kind of experiments, one should be extremely careful about the possible magnitude of baryonic effects at small scales, because, although the results obtained with the wrong model may appear as a good fit, the corresponding bias can be in fact as high as many $\sigma$. Also, the recovered parameters from the wrong model (DMO) move away from the true value with increasing $\ell_{max}$. This suggests a potential test for a given model, the cosmological parameter space should not move significantly when analysing up to different scales, the difference should only be seen in the variance of the parameters and not in its mean value.

\subsection{$M_{\rm crit}$ parameter}

The only free parameter in our BAR model, $M_{\rm crit}$, regulates the amount of gas inside the halo as a function of halo mass. We explore the consequences of what we believe to be a realistic model ($M_{\rm crit} = 10^{13} h^{-1}M_{\bigodot}$) in details considering nine different $\ell_{max}$. At $\ell_{max}=1000$, there is hardly any constraint drawn from weak lensing on this parameter, but as we increase $\ell_{max}$, baryonic physics become more and more important and thus constraints can be put on $M_{\rm crit}$. In fact, the constraints on this parameter increase rapidly from 15\% at $\ell_{max}=1000$ to 1-2\% at $\ell_{max}=4000$. After this, no significant improvement on the constraints can be gained on this parameter. The variance of $M_{\rm crit}$ becomes constant after nearly $\ell_{max}=8000$, which is what happens for the other cosmological parameters. So, with this kind of weak lensing experiment, $M_{\rm crit}$ (or $\log(M_{\rm crit})$) could be constrained up to 1-2\%, which is quite impressive.

\label{subsec:goodness}

\subsection{Non-Gaussian covariance vs baryonic corrections}
\label{sec:NG}
Being able to extract cosmological information from clustering data down to a few percent accuracy can be considered very optimistic. It can be jeopardized by many unresolved issues. The two most important issues are (i) baryonic physics at small scales, and (ii) non-Gaussian effects in the covariance matrix of the power spectrum. These two issues can be quantified in projected weak-lensing statistics, like the shear power spectrum. In this work, we primarily talk about the effect of baryonic physics at small scales on the shear power spectrum and its cosmological implications. However, we ignore the effect of non-Gaussianity (NG) on the covariance matrix. 

The NG contribution to the covariance becomes more important at small scales, like baryonic physics \citep{2009MNRAS.395.2065T, 2013PhRvD..87l3504T}. Now the question is, which one is more important to deal with and which one appears first when going towards smaller scales? This question does not have a very straightforward answer. Ignoring both  of these contributions may result in highly biased cosmological parameters estimations.

\cite{2012PhRvD..86h3504Y} (figure 9, right panel) shows the constraints on the amplitude of fluctuations ($\sigma_8$) as we go to smaller scales. If one considers only Gaussian errors, the constraints continue to improve until the instrumental shot noise kicks in. However, NG contribution are likely to dominate over Gaussian errors after $\ell=700$. But we cannot directly compare to this plots as the constraints depend on many other details. We can still compare the ratios of the NG and Gaussian contributions. At $\ell_{max}=10000$, the NG covariance is six times the Gaussian covariance. On the other hand, in figure \ref{fig:bias} the bias in cosmology becomes close to 10$\sigma$ for $\sigma_8$ at $\ell=10000$. This means that the NG corrections are sub-dominant than the baryonic effects. However, our analogy is very hand-wavy and requires further study.

\subsection{The ideal configuration}

We explore the baryonic effects on the cosmological parameter estimation and found big bias in cosmological parameters if the analysis include $\ell>4000$. After this limit, the cosmological parameters start to become biased and mislead the constraints. However, the constraints keep improving up to $\ell=10000$. So the question arises, what is the ideal configuration to perform weak-lensing power spectrum analysis to put useful constraints on cosmology with Euclid-like surveys? 

We explore this answer in our analysis and stated our results in the previous sections. To summarize, the ideal configuration is to go as high as $\ell=8000$, including baryonic physics and marginalize over the baryonic parameters, in our case $M_{\rm crit}$. In this configuration, one can find unbiased estimates of the cosmological parameters. Having the unbiased estimates, we can also constrain the cosmological parameter space with much better accuracy than before. In this configuration, $\Omega_m$ and $\sigma_8$ can be estimated with nearly 2\% and 0.5\% respectively. The variance of the two parameters defining the redshift evolution of the equation of state of dark energy, $w_0$ and $w_a$ are 0.07 and 0.15 respectively. Along with cosmological parameters, the baryonic parameter $M_{\rm crit}$ can also be estimated to very high accuracy, as good as 1-2\%.

When dealing with real clustering datasets, we are also able to use independent constraints on the baryonic parameters, such as abundance matching data and/or X-ray data on individual halos, providing a solid understanding of the overall signal and the underlying baryonic effects.

%===============================================================================
\section{Acknowledgement}
I.M. would like to thank Prasenjit Saha, Uros Seljak and Ravi Sheth for useful discussions about the topic and their suggestions.
%===============================================================================\
\bibliographystyle{mn2e}
%===============================================================================
\def\apj{ApJ}
\def\apjl{ApJL}
\def\apjs{ApJS}
\def\aj{AJ}
\def\prd{PRD}
\def\mnras{MNRAS}
\def\aap{A\&A}
\def\physrep{PhysicsReports}
\def\nat{Nature}
\def\araa{ARAA}
%===============================================================================
\bibliography{manuscript.bib}
%===============================================================================
\end{document}